\documentclass[fleqn,usenatbib]{mnras}

\usepackage{newtxtext,newtxmath}
\usepackage{tablefootnote}
\usepackage[T1]{fontenc}

\DeclareRobustCommand{\VAN}[3]{#2}
\let\VANthebibliography\thebibliography
\def\thebibliography{\DeclareRobustCommand{\VAN}[3]{##3}\VANthebibliography}

\usepackage{graphicx}	% Including figure files
\usepackage{amsmath}	% Advanced maths commands
\usepackage{comment}	
\usepackage{orcidlink}
\title[Semi-analytic modelling of Pop. III stars]{Semi-analytic modelling of Pop. III star formation and metallicity evolution - I. Impact on the UV luminosity functions at $z = 9-16$}

\author[E. M. Ventura et al.]{Emanuele M. Ventura$^{\orcidlink{0000-0003-3502-4929}}$$^{1,2}$\thanks{E-mail: eventura@student.unimelb.edu.au},
Yuxiang Qin$^{\orcidlink{0000-0002-4314-1810}}$$^{1,2}$,
Sreedhar Balu$^{\orcidlink{0000-0002-5281-5151}}$$^{1,2}$
and J. Stuart B. Wyithe$^{\orcidlink{0000-0001-7956-9758}}$$^{1,2,3}$
\\
$^{1}$School of Physics, University of Melbourne, Parkville, Victoria, Australia\\
$^{2}$ARC Centre of Excellence for All Sky Astrophysics in 3 Dimensions (ASTRO 3D)\\
$^{3}$Research School of Astronomy and Astrophysics, Australian National University, Canberra, ACT 2611, Australia\\
}

\date{Accepted XXX. Received YYY; in original form ZZZ}

\pubyear{2023}

\begin{document}
\label{firstpage}
\pagerange{\pageref{firstpage}--\pageref{lastpage}}
\maketitle

\begin{abstract}
We implemented Population III (Pop. III) star formation in mini-halos within the \textsc{meraxes} semi-analytic galaxy formation and reionisation model, run on top of a N-body simulation with $L = 10 h^{-1}$ cMpc with 2048$^3$ particles resolving all dark matter halos down to the mini-halos ($\sim 10^5 M_\odot$). Our modelling includes the chemical evolution of the IGM, with metals released through supernova-driven bubbles that expand according to the Sedov-Taylor model. We found that SN-driven metal bubbles are generally small, with radii typically of 150 ckpc at $z = 6$. Hence, the majority of the first galaxies are likely enriched by their own star formation. However, as reionization progresses, the feedback effects from the UV background become more pronounced, leading to a halt in star formation in low-mass galaxies, after which external chemical enrichment becomes more relevant. We explore the sensitivity of the star formation rate density and stellar mass functions on the unknown values of free parameters. We also discuss the observability of Pop. III dominated systems with JWST, finding that the inclusion of Pop. III galaxies can have a significant effect on the total UV luminosity function at $z = 12 - 16$. Our results support the idea that the excess of bright galaxies detected with JWST might be explained by the presence of bright top-heavy Pop. III dominated galaxies without requiring an increased star formation efficiency. %227 words
\end{abstract}


%%%%%%%%%%%%%%%%%%%%%%%%%%%%%%%%%%%%%%%%%%%%%%%%%%

%%%%%%%%%%%%%%%%% BODY OF PAPER %%%%%%%%%%%%%%%%%%

\section{Introduction}

The first episodes of star formation likely occurred at redshift 30-40 inside low-mass (10$^{5-7}$M$_\odot$) halos with a pristine chemical composition (metal-free or extremely metal-poor). These environments were responsible for the chemical enrichment of the intergalactic medium (IGM) leading to the Pop. III/Pop II transition at redshift 15-10. However, due to the low mass and the high-$z$, the study of Pop. III star formation in mini-halos is challenging both in terms of observations and simulations (see \citealt{Klessen2023} for a recent and comprehensive review).\\ Even with the launch of JWST, we do not have confirmed observations of Pop III stars and only a few potential candidates (\citealt{Welch2022,Maiolino2023}). As pointed out by \citet{Trussler2023}, a direct detection of a Pop. III galaxies will be extremely challenging as, in the best-case scenario, hundreds of hours of integration time are needed in order to detect an unlensed Pop. III system with 5$\sigma$ confidence. 
%A complementary tool to direct observation could be the redshifted 21cm global signal whose depth might be determined by the radiation emitted from Pop. III stars and early accreting black holes during the Cosmic Dawn (among the most recent works see \citealt{Mebane2018,Mirocha2018,Mebane2020,Munoz2022,GesseyJones2022,Magg2022,Ventura2023,Hegde2023}). So far, there is no detection confirmed of the 21cm redshifted line, however, a large number of facilities aiming to the observation of the 21cm global signal and power spectrum are already operating (LOFAR \citet{Mertens2020}, MWA \citet{Trott2020}, PRIZM \citet{Philip2019}, SARAS3 \citet{Singh2022}, HERA \citet{DeBoer2017}) or becoming operative in this decade (e.g. NenuFAR \citet{Mertens2021}, REACH \citet{Cumner2022}, SKA \citet{Koopmans2015} etc.)\\ %Double check a few references if there is more recent stuff
There are a number of Pop. III star formation simulations at different scales. 
%The smaller scales (L $< 1$ comoving Mpc) are mostly used to constrain the detailed physics that regulates the formation of individual stars. 
Hydrodynamical simulations that follow the cooling of gas in a single pristine or very metal-poor cloud suggest that the inefficient cooling due to the lack of metals might favour an initial mass function (IMF) that is shifted to larger masses \citep[e.g.][]{Bromm1999, Hirano2014, Stacy2016, Chon2021}{}{}; however, there is no general consensus (e.g. \citealt{Wollenberg2020,Jaura2022,Prole2022} predict IMF shifted to lower masses). 
Simulations at larger scales ($\geq$1 cMpc) are instead used to follow the global evolution of Pop. III stars with redshift and their impact on the cosmic metal enrichment and reionization. In the last decade, many of these simulations have been performed both with hydrodynamical (\citealt{Pallottini2014,Xu2016,Sarmento2018,Jaacks2018,Skinner2020,Schauer2021,Venditti2023,Magnus2023,Yajima2023}) and semi-analytical codes (\citealt{Trenti2009,Crosby2013,Mebane2018,Visbal2018,Visbal2020,Trinca2022,Magg2022,Hegde2023,Feathers2023}). %Looking at larger scales however, both approaches encounter a resolution problem. 
%To study the first episodes of star formation in mini-halos we need to have a high-resolution in order to account in a self-consistent way all the small-scale physics such as the stellar feedback. At the same time we also need to consider a volume large enough in order to have a statistically significant sample of the Universe (in order to form massive halos we need a cube of L $\geq 100$ cMpc). However, to satisfy both requirements of large volumes and high-resolution is impossible and thus all the semi-analytical and hydrodynamical simulations have to make compromise in one of the two directions. \\
In this work, we chose to use the semi-analytical model of galaxy formation \textsc{Meraxes} (\citealt{Mutch2016}, M16 hereafter) within a simulation that allows us to resolve all the mini-halos down to a few 10$^5$M$_\odot$ in a simulation of L = 10 Mpc $h^{-1}$. 
%We emphasize that this work is only the first step of the more ambitious project of capturing the small scale physics in a big box. As so, in this paper we ran \textsc{Meraxes} on top of a N-body simulation of L = 10 Mpc. We are well aware that 
The size of this simulation is not large enough to be representative of the Universe (we will miss the most massive galaxies). However, the scales are large enough to investigate the impact of the main physical processes on the Pop. III star formation.
We incorporated a number of new physical processes that are relevant to Pop. III star formation in mini-halos including: \textit{(i)} molecular hydrogen cooling functions for mini-halos, \textit{(ii)} baryon-dark matter streaming velocities, \textit{(iii)} photo-dissociation of H$_2$ molecules from the Lyman-Werner background and \textit{(iv)} the chemical evolution of the intergalactic medium. This latter effect has been implemented assuming that metals are released through supernova explosions and within a bubble that expands accordingly to the Sedov-Taylor model (the approach is very similar to the analytical calculation shown in \citealt{FL2003}). We found that these bubbles are generally small and that they roughly agree with the previous estimate by \citet{Trenti2009}, with typical radii of 150 ckpc at $z = 6$. The implementation of both internal and external metal enrichment allows us to understand whether a galaxy will form Pop. III or Pop. II stars and thus quantify the impact of Pop. III galaxies on the total luminosity function at $z \geq 5$.
%To do that we implemented Pop. III spectral energy distributions (SEDs) from \citet{Raiter2010} and we assumed that the Pop. III star formation occurs in an instantaneous burst at a random $\Delta t$ from the end of the snapshot. Considering an instantaneous rather than a continuous star formation makes some of the Pop. III dominated systems significantly brighter, hence at high-$z$ the brightest systems are likely to host Pop. III stars. 
%We found that, given their lower total mass and shorter lifetimes, Pop. III dominated galaxies are extremely challenging to detect even with JWST deep surveys that go up to 30.0 apparent mag. The chance of observing these systems increases if we consider a high Pop. III star formation efficiency and a log-normal IMF with a characteristic mass of $70 M_\odot$. We also  \\ %You might add/update this part here
This paper is structured as follows:\\
In Section \ref{sec:Model}, we present the new physics implemented in \textsc{Meraxes}. In Section \ref{sec:global}, we show the main global properties of the Pop. III star formation across cosmic history discussing which parameters have a stronger impact on the global star formation history and on the metallicity of the IGM. In Section \ref{sec:observability} we discuss the SED evolution of Pop. III galaxies and the UV luminosity functions at $z = 9 - 16$ for different Pop. III IMFs. Finally, we discuss our main results and conclusions in Section \ref{sec:conclusions}.

\section{Pop. III star formation in MERAXES}
\label{sec:Model}
The semi-analytical model \textsc{meraxes} was first developed by \citet{Mutch2016} (hereafter M16), \citet{Qin2017,Qiu2019} in order to study galaxy formation and growth through the Epoch of Reionization. 
%and it undergone through several updates. Compared to the original version showed in \citet{Mutch2016} the main ones involved detailed prescriptions for AGN feedback (\citealt{Qin2017}) and for supernova feedback and chemical enrichment inside each galaxy (\citealt{Qiu2019}). Despite the variety of physical processes included in 
\textsc{meraxes} assumes that all the galaxies form in a previously chemically enriched Universe and inside atomic cooling halos. Such an approximation does not allow us to study the first episodes of star formation that mostly occurred in mini-halos when the Universe did not have any metals. The version of \textsc{meraxes} presented in this work allows us to compute the physics of the first episodes of star formation from the initial molecular cooling of the gas to the external metal enrichment from the supernova feedback. As in the previous work, \textsc{meraxes} is coupled to the reionization so that all the radiative backgrounds are computed in a self-consistent way from the galaxy properties. This has been done by implementing a modified version of 21cm\textsc{fast} (\citealt{Mesinger2011}) that includes the local ionizing UV background from \citet{Sobacchi2014} and the X-ray heating (\citealt{Balu2023}). In this version, we also included the Lyman-Werner background. 
%in the same version presented in \citet{Qin2020} adapted for \textsc{meraxes} (see section \ref{sec:LW} for the details in the implementation).

\subsection{High-Resolution N-body simulation}
\label{sec:L10} 
The updates to \textsc{meraxes} for Pop. III necessitate high mass and spatial resolution. For this purpose, we introduce L10\_N2048 (hereafter L10) from the \textit{Genesis} suite of dark matter only \textit{N}-body simulations.
L10 is a periodic cubical simulation of side 10 $h^{-1}$ Mpc and consists of 2048$^3$ dark matter particles of mass m$_{\rm p}$ = 9.935 $\times 10^3 h^{-1} M_\odot$ resulting in a halo mass resolution of $\sim$ 3.18 $\times 10^5 h^{-1}M_\odot$ (based on a minimum of 32 particles per halo). The simulation, run using the \textsc{SWIFT} \citep{Schaller2018} cosmological code, evolves these dark matter particles from $z = 99$ down to $z = 5$. The halos were identified using the friends-of-friends phase space  halo-finder \textsc{VELOCIraptor} \citep{VR} and the merger trees were constructed using \textsc{TREEFROG} \citep{treefrog}. We note that the mass resolution achieved in this simulation allows us to resolve all the mini-halos below $z \sim 30$ and is the highest resolution on which \textsc{meraxes} has been run. The output trees of the N-body simulation used in this work are available on Zenodo at \citet{balu2024}.

\subsection{Molecular cooling}
\label{sec:cooling}
The first process in order to enable star formation is the cooling of the gas. For dark matter halos with $T_{\rm vir} \geq 10^4$ K, the main coolant is the atomic hydrogen, while in mini-halos ($10^3$ K $\leq T_{\rm vir} < 10^4$ K), the cooling occurs via roto-vibrational transitions of molecular hydrogen \citep[e.g.][]{Tegmark1997}. For the details on how the cooling of the gas is implemented in \textsc{meraxes}, we refer the reader to M16; in this section, we only highlight the main differences between the atomic and the molecular cooling.\\
As in M16, we compute the ratio of the specific thermal energy to the cooling rate per unit volume:
\begin{equation}
    t_{\rm cool}(r) = \frac{1.5\Bar{\mu} m_p kT}{\rho_{\rm hot}(r)\Lambda(T,f_{H_2})}
\label{eq:MC}
\end{equation}
where $\rho_{\rm hot}(r)$ is the gas density profile (we assume a singular isothermal sphere), k is the Boltzmann constant, and T is the temperature of the gas (which we set to be the virial temperature of the halo). M16 included only atomic cooling halos. Thus they set the mean molecular weight $\Bar{\mu} = 0.59$ assuming a fully ionized gas
%mean particle mass $\Bar{\mu}{\rm m}_p = 9.868 \times 10^{-25}$g assuming a fully ionized gas 
and the cooling function $\Lambda (T)$ in units of erg cm$^{-3}$ s$^{-1}$ from \citet{Sutherland1993}. For mini-halos we instead set $\Bar{\mu} = 1.22$ (fully neutral gas)
%\Bar{\mu}m_{\rm p} = 2.041 \times 10^{-24}$g 
and implement the molecular hydrogen cooling functions $\Lambda (n_{\rm H}, T)$ from \citet{Galli1998}. This choice is valid as long as we assume that the cooling inside mini-halos occurs only due to H$_2$. This may not be valid as, if a mini-halo is chemically enriched with metals by a nearby halo, metals are much more effective in the cooling of the gas \citep[e.g.][]{Nebrin2023}{}{}. However, as we will show in section \ref{sec:global}, this enrichment is almost ineffective at $z > 10$. The molecular cooling function $\Lambda$ depends on the gas density of the halo (which can be directly computed for each galaxy assuming an isothermal sphere) and on the molecular hydrogen fraction $f_{H_2}$. For the latter we assumed a fixed value of 0.1\%, which is consistent with results from \citet{Nebrin2023} for halos of T$_{\rm vir}\simeq5\times10^3$K at $z = 15-20$\footnote{The exact value of $f_{\rm H_2}$ depends on the virial mass of the halo, redshift and the time available for H$_2$ formation in the minihalo (which is likely to be a multiple of the free fall time). To compute the exact value of $f_{\rm H_2}$ is beyond the aim of this work, hence we chose an average value.}. As in M16, the cooling time is used to find a cooling radius that determines the amount of gas that is cooled down.

%PREVIOUS VERSION: This cooling function depends on the hydrogen number density $n_{\rm H}$ for which we adopted the fitting function from \citet{Trenti2009b} whose result is roughly in agreement with the gas densities in mini-halos found in \citet{OLeary2012}. As in M16, the cooling time is used to find a cooling radius that determines the amount of gas that is cooled down.
%There are a few things that should improve but overall is ok. Atm leave it like that. Possible improvements are the adding of the streaming velocity (very easy to do and more updated fitting functions).
\subsection{Streaming velocities}
\label{sec:SV}
As a first approximation, gas in a mini-halo can start to cool down the gas once it reaches the virial temperature of $\simeq 10^3$ K. From \citet{Barkana2001}, adopting $\mu = 1.22$ for a fully neutral gas, this requirement would correspond to a minimum virial mass of (\citealt{Visbal2015}):
\begin{equation}
    M_{\rm min, H_2} = 2.5 \times 10^5\bigg(\frac{26}{1+z}\bigg)M_\odot.
    \label{eq: MminH2}
\end{equation}   
However, there are a number of effects that can decrease the amount of molecular hydrogen present in the halo, reducing the cooling efficiency and ultimately increasing the minimum virial mass of a halo capable of cooling (see \citealt{Nebrin2023} for an extensive discussion and comparison between results found in different simulations.)\\
A non-radiative process that can delay the gas cooling in very low-mass halos is the streaming velocity between baryons and dark matter (\citealt{Tseliakhovich2010}). This effect is a consequence of the different decoupling of baryons and dark matter particles from photons that results in a velocity difference between the two species of $v_{\rm bc}$ that can be expressed as a constant multiple of the root-mean-square (rms) streaming velocity $\sigma_{\rm bc} = 30 \times (1+z)/1000$ km s$^{-1}$. The presence of a relative motion between baryons and dark matter particles will make it harder for baryons to fall into the potential wells of dark matter halos, delaying the accretion and, hence, the cooling of the gas in mini-halos. The main outcome of this physical process is to delay the very first episodes of star formation around $z \sim 40$. Throughout this work, we implemented the effect of the streaming velocities as per other semi-analytical models based on a fitting function found by \citet{Fialkov2012}, which is calibrated to reproduce the results of the hydrodynamical simulations of \citet{Greif2011} and \citet{Stacy2011}:

\begin{equation}
    V_{\rm cool, H_2} = (a^2 + (bv_{\rm bc})^2)^{0.5},
    \label{eq:VminStream}
\end{equation}
where a = 3.714 km s$^{-1}$ and b = 4.015 km s$^{-1}$. Equation \ref{eq:VminStream} provides the minimum circular velocity that a halo needs to have in order to have enough H$_2$ to cool down the gas. V$_{\rm cool, H_2}$ can be easily converted into a virial mass M$_{\rm cool, H_2}$ using the formula in \citet{Barkana2001}. In this work we fixed $v_{\rm bc}(z) = \sigma_{\rm bc}(z)$ and we assumed this value for the entire box.\footnote{This assumption is not entirely accurate as the streaming velocities are roughly constant up to a scale of a few Mpc while our box is larger than 10 Mpc. However, the effect of streaming velocities is dominant only at $z \geq 30$ before the first stars form, and the Lyman-Werner background is built up.}  
%However, we argue that the effect of streaming velocities in reducing the H$_2$ cooling efficiency is dominant only at $z \geq 30$ before the first stars form (its main effect is to delay the very first episodes of star formation around $z \sim 40$). Once the first mini-halos are forming stars, the effect of the streaming velocities becomes widely subdominant compared to the photo-dissociation of the Lyman-Werner background (see following section). In a future work focused on different dark matter models we will explore the impact of different values of $v_{\rm bc}$ and of a non-uniform streaming velocity field.

%and using the correction factor of 0.75 suggested by \citet{Fernandez2014} 1.6 \times 10^6 ((1+z) / 10)^{-1.5}

\subsection{LW background}
\label{sec:LW}

A self-consistent model of star formation in mini-halos must consider the photo-dissociation of H$_2$ from UV photons in the Lyman-Werner (LW) band ([11.2 - 13.6] eV). LW photons destroy H$_2$ and thus prevent mini-halos gas from cooling (\citealt{Haiman2006}). We implement this effect by changing the minimum mass for molecular cooling using the fitting from \citet{Visbal2015}:
\begin{equation}
    M_{\rm crit,MC} = M_{\rm cool, H_2}[1 + 22.87 \times J^{0.47}_{\rm LW}].
    \label{eq:McritMC}
\end{equation}
This critical mass only applies to minihalos which are below the atomic cooling halo mass threshold.
In the absence of an LW background, this equation simply gives M$_{\rm crit,MC}$ as defined in Section \ref{sec:SV}. J$_{\rm LW}$ is the LW flux in units of 10$^{-21}$ erg s$^{-1}$ cm$^{-2}$ Hz$^{-1}$ sr$^{-2}$ that reaches the minihalo. LW photons have a mean free path of $\sim 100$Mpc, so each minihalo will be affected even by distant galaxies that formed at higher redshift. Thus, we modelled the LW background by integrating contributions across the cosmic history (see a similar approach in \citealt{Qin2020} which is briefly summarized below.) \\
%adopting the same approach presented in \citet{Qin2020} and hereafter briefly summarized.\\ 
As for all the radiative backgrounds in \textsc{meraxes}, we follow an excursion-set formalism (\citealt{Furlanetto2004}) which counts the number of photons in a certain band in spheres of radius R centred at location and redshift (\textsc{x},z). We decrease the radius down to the size of the cell R$_{\rm cell}$.
At each of these locations, we compute the LW emissivity $\epsilon_{\rm LW}(\textbf{x}, z')$ using the spectral energy distributions from \citet{Barkana2005}. These give $\frac{dn}{d\nu}$: the number of photons per solar mass per unit frequency for both Pop III and Pop II stars. We assume that LW photons are absorbed only at resonant frequencies (thus we are neglecting any absorption from H$_2$ in the IGM) where $\nu_{\rm n} = \nu_{\rm LL}(1 - n^{-2})$ and $\nu_{\rm LL} = 3.29 \times 10^6$ GHz is the Lyman limit frequency. Under these approximations, we can compute the LW emissivity smoothed over R at redshift $z$ and location $x$ for sources emitted at redshift $z'$ as: 
\begin{equation}
    \epsilon_{\rm LW}(\textbf{x},z'|z,R) = \sum_{\rm {i = III, II}}{\rm{SFRD}}^i\sum_{n = 2}^{n_{\rm max}(z)}\int_{\rm max(\nu'_n,\nu_{LW})}^{\nu_n + 1}\frac{dn^i}{d\nu''}hd\nu''
\label{eq:epsLW}
\end{equation}
In the equation above, we directly link the star formation rate density (SFRD) for both Pop. III and Pop. II stars (which we can compute from \textsc{meraxes}) to the LW emissivity summed over the Lyman series. This sum accounts for the resonances in the Lyman series (Lyman-n photons will be absorbed and re-emitted as a Lyman$-\alpha$ photon). Given the large mean free path of LW photons, we must also account for distant galaxies at higher redshift $z' > z$ with the redshifted spectrum $\nu' = \nu\frac{1+z'}{1+z}$. The sum of these two effects causes the peculiar shape of the LW spectrum (see Fig. 2 in \citealt{Qin2020}). Following \citet{Barkana2005} we account for all the Lyman resonances with $n \leq 23$. Under these approximations we can find the maximum redshift $z_{\rm max}$ from which a LW photon can reach $z$:
\begin{equation}
    z_{\rm max} + 1 = (z+1)\frac{1-(n+1)^{-2}}{1-n^{-2}}.
\end{equation}
Finally, we need to convert the emissivity into a flux following (\citealt{Qin2020}):
\begin{equation}
    J_{\rm LW}(\textbf{x},z|R) = \frac{(z+1)^3}{4\pi}\int_z^{\infty}dz'\frac{cdt}{dz'}\epsilon_{\rm LW}.
    \label{JLW}
\end{equation}
The approach described above differs from the one in 21cm\textsc{fast} only for the computation of the SFRD. In \citet{Qin2020}, the SFRD is estimated from the density field and the collapsed fraction, while in this work, it is computed directly from \textsc{meraxes}, which tracks the formation of each galaxy and its entire star formation history.\footnote{In this work, we are not accounting for the self-shielding of H$_2$, which acts against the H$_2$ photo-dissociation and thus might enhance the molecular cooling in mini-halos (\citealt{Hartwig2015, Schauer2021, Chiaki2023}) increasing the Pop. III SFRD (\citealt{Feathers2023}). \citet{Kulkarni2021} provided a fitting formula for M$_{\rm crit, MC}$ accounting for the H$_2$ self-shielding. Since this affects only the LW feedback, the combined effect of the LW background and the streaming velocities in \citet{Kulkarni2021} is not multiplicative as assumed in Eq. \ref{eq:McritMC}}.\\ 
%However, given that the volume of our simulation (L = 10 cMpc) is much smaller than the mean-free path of LW photons, each halo is losing the contribution of the most distant sources, thus neglecting the self-shielding should partially counterbalance the loss of the LW background from distant galaxies.}\\

Together with the LW background, the main radiative feedback that suppresses star formation is the UV photo-ionization (the ionizing UV background inhibits gas accretion in low-mass halos \citealt{Sobacchi2013}). We used the same prescriptions as in M16, which consists of reducing the baryon content through a baryon fraction modifier f$_{\rm mod}$ (that depends on the local ionizing rate and the time when the nearby IGM becomes ionized) that stops the gas infall. In difference from the LW feedback, the UV photo-ionization is relevant only during reionization ($z \leq 10$), and it also affects atomic cooling halos as massive as $\sim 10^{9.5} M_\odot$.
%In this work we are not accounting for the self-shielding of H$_2$ which acts against the H$_2$ photo-dissociation and thus might enhance the molecular cooling in mini-halos (\citealt{Hartwig2015, Chiaki2023}). However, given that the volume of our simulation (L = 10 cMpc) is much smaller than the mean-free path of LW photons, each halo is losing the contribution of the most distant sources, thus neglecting the self-shielding should partially counterbalance the loss of the LW background from distant galaxies. 
%In future work we will apply this model to a larger box (L $\geq$ 100 cMpc) including the self-shielding.  

\subsection{Metal Evolution}
\label{sec:Metals}
Once stars start to form in the Universe, they also explode as supernovae, releasing metals. Most of these will stay inside the same galaxy, contributing to the chemical enrichment of the galaxy itself. This process is commonly known as "genetic" enrichment and is likely to be the main mechanism of chemical enrichment of the Universe. However, some of the metals escape their parent galaxy. In this case, they will pollute the nearby IGM, and if later a galaxy forms in a region where the IGM was enriched, the new galaxy will be pre-enriched with metals. This latter mechanism is referred to as "external" metal enrichment \citep[e.g.][]{Pallottini2014,Smith2015,Hartwig2018,Visbal2020,Yamaguchi2023}{}{}. Keeping track of the metallicity evolution of the Universe is crucial in order to put constraints on when the Pop. III/II transition occurred. \\
We account for both processes, and we are able to follow the metallicity evolution of the IGM. Firstly, we choose a critical metallicity $Z_{\rm crit} = 10^{-4}Z_\odot$ as the threshold value below which a galaxy will form Pop. III stars. All new galaxies, unless externally polluted, will accrete pristine gas (without any metals) onto the hot gas reservoir. This gas, once it cools, will provide the reservoir for the star formation. Hence, a galaxy that is not externally polluted will always form Pop. III stars for the first time. At each snapshot, we compute the metallicity of the cold gas reservoir from the amount of metals released by earlier supernovae and if this is higher than $Z_{\rm crit}$, the galaxy will form Pop. II stars, otherwise it will form Pop. III.\\  
% OLD: Firstly, we choose a critical metallicity $Z_{\rm crit} = 10^{-4}Z_\odot$ as the threshold value below which a galaxy will form Pop. III stars. All the new galaxies, unless externally polluted, are assumed to form with no metals, hence when they will form stars for the first time in their history, they will form Pop. III stars. Any following star formation episode will be Pop. II, hence galaxies can form Pop. III stars only once in their life. This assumption is equivalent to say that the internal enrichment is instantaneous. Even if it is not probably the case this approximation is well motivated in \textsc{meraxes} given that the time separation between two consecutive snapshots is $\Delta t \geq 2$ Myr, and each galaxy can have only one stellar population.\\ 
While internal enrichment is the main mechanism that drives the Pop.III/II transition, some galaxies can be externally enriched through supernova winds originating in a nearby galaxy. Once several supernovae in a galaxy go off, they will form a "super-bubble" that will expand outside the galaxy escaping the binding energy of the dark matter halo. We followed the expansion in time of this "metal bubble" using the analytic approximation in \citet{Dijkstra2014}:

\begin{equation}
    r_{\rm bubble}(t) = \bigg(\frac{\Delta E_{\rm SN}}{m_{\rm p}n_{\rm gas}}\bigg)^{0.2}t^{0.4}.
    \label{eq:Rbubble}
\end{equation}
All quantities that appear in this equation are computed in \textsc{meraxes}. $\Delta E_{\rm SN}$ is the total supernova energy released at a certain snapshot, and it is computed using the stellar mass that goes supernova in that snapshot multiplied by the supernova explosion energy (see Eq. 11 in M16). n$_{\rm gas}$ is the number density of the gas to which the bubble has expanded (while the bubble is smaller than the virial radius, this is the gas density of the galaxy; otherwise, the density of the IGM), and t is the time since the explosion occurred. We assume that all the supernova events will occur in the middle of the snapshot\footnote{This assumption is not self-consistent with the luminosity computation of Pop. III galaxies, which assumes stochastic Pop. III star formation (see section \ref{sec:PopIII_spectra}).}. Note that since \textsc{meraxes} accounts for both contemporaneous and delayed supernova feedback (see more in M16), each galaxy has several bubbles associated with the same star formation episode. However, we consider only the largest of these bubbles so that each galaxy has only one associated bubble. Having calculated the bubble size, we can predict if a nearby galaxy will accrete pristine or enriched gas. In order to reduce the computational cost, we avoid computing the distance between all the pairs of galaxies and instead use the grid-based approach outlined below:
\begin{itemize}
    \item Build a high-resolution grid with $N_{\rm grid}^3$ cells smoothing the matter density grid from the N-body simulation. The size of the cell is chosen so that the volume of the cell is similar to the volume of the typical bubble at the end of the simulation ($R_{\rm ave, bubble} \simeq 100$ckpc, see Fig. \ref{fig:BubbleDF}). For our box, this requirement is satisfied when $N_{\rm grid} = 128$. This ensures that the metal bubble will not overflow outside the pixel where it originates. 
    \item For each cell i, we compute the average metallicity $Z_{\rm IGM,i}$ as the ratio between the sum of the metals ejected by all galaxies inside the cell and the total gas in the cell:
\begin{equation}
Z_{\rm IGM,i} = \sum_j^N\frac{M_{\rm MetalsEj,j}}{M_{\rm IGM,i} + M_{\rm EjGas,j} - M_{\rm gas,j}}.  
\label{eq:ZIGM}
\end{equation}
    Here $M_{\rm IGM,i} = \rho_{\rm crit}\Omega_m(1+\delta_i)f_bd^3$ is the total baryonic mass inside each cell which depends on the average matter overdensity of the cell $\delta_i$ and the cell size $d$. $M_{\rm EjGas}$ and $M_{\rm gas}$ are respectively the total mass of the ejected gas and the bounded gas of each galaxy. We only let the galaxies with $r_{\rm bubble} \geq 3 \times R_{\rm vir}$ contribute to $M_{\rm MetalsEj,j}$ and $M_{\rm EjGas,j}$. For the computation of the metals ejected from Pop. II galaxies, we refer the reader to M16 and \citet{Qiu2019} and for Pop. III galaxies to Section \ref{sec:PopIII}.
    \item For each cell i, we compute the volume fraction filled with metals (the metal filling fraction), summing the volume of the largest bubble surrounding each galaxy
    %all the metal bubbles 
    within the pixel and dividing by the cell volume. As above, we account only for those galaxies with $r_{\rm bubble} \geq 3 \times R_{\rm vir}$. This choice reflects the fact that a central galaxy needs to be surrounded by a bubble at least as large as its virial radius in order to pollute a nearby galaxy. The factor of 3 has been chosen given that the distance between two central galaxies in our simulation is always larger than three times the virial radius of the biggest galaxy. Hence, the metal filling factor for the cell i is $Q_{\rm metal,i} = \frac{\sum_j^NV_{\rm bubble,j}}{V_i}$. 
    \item At the beginning of each snapshot, we assign the probability p for external metal enrichment to each galaxy inside the cell i. This probability will be given by the metal filling factor $Q_i$.
    This choice assumes that the galaxies are randomly distributed within each pixel. This assumption holds only if we use a high-resolution grid where the pixel size is small enough such that clustering effects are negligible. 
    %We refer the reader to Appendix \ref{sec:AppendixA} for a comparison between the adopted model and another model with a low-resolution grid that accounts for the clustering effect.
    \item We assign a random number $m$ between 0 and 1 to each newly formed galaxy, and at each snapshot, we compare this random number to $p$. When the condition $m \leq p$ is satisfied, we label that galaxy as externally enriched, and it accretes enriched gas with a metallicity equal to the average metallicity of the cell that belongs to ($Z_{\rm IGM,i}$). Furthermore, when a galaxy experiences a star formation episode, we enforce $p = 1$ 
    %and assume it will accrete gas with a metallicity computed from its ejected reservoir 
    (in this latter case, we know that this galaxy will be inside its own metal bubble and thus cannot accrete pristine gas). With this latter condition and internal enrichment, we effectively stop Pop. III star formation inside a galaxy after the first supernova episode.
\end{itemize}

\begin{figure*}
    \centering
    \includegraphics[width=\textwidth]{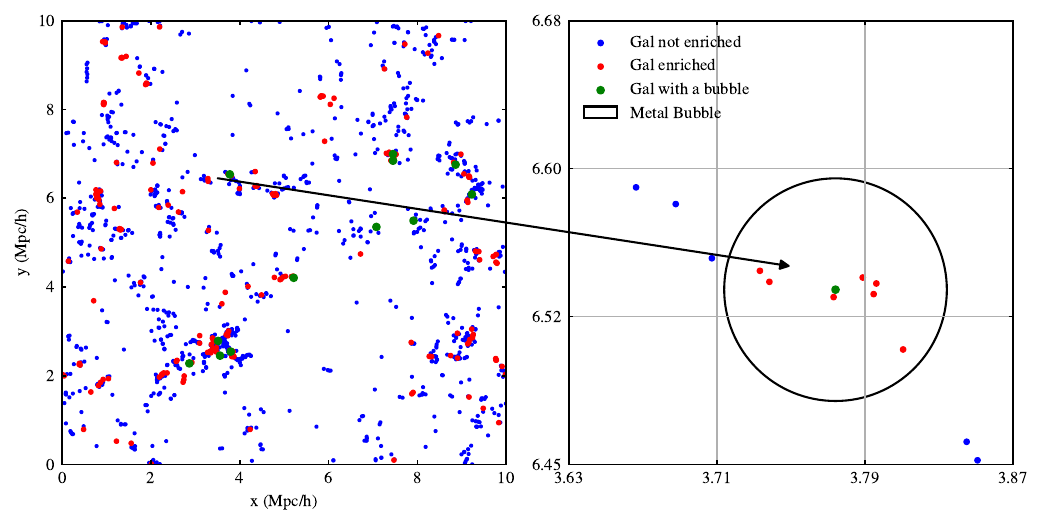}
    \caption{\textbf{left:} 2D slice of the L10 box at $z = 10$ with the position of all the galaxies with $M_{\rm vir} \geq 10^6 M_{\odot}$ (blue dots) highlighting the ones that are externally metal enriched (red dots). The green dots are galaxies with an associated metal bubble larger than $0.05$ $h^{-1}$ cMpc. The slice is 10 $h^{-1}$ Mpc on a side and $\sim0.078 h^{-1}$Mpc thick. \textbf{right:} zoom-in of a single pixel (0.24 $h^{-1}$ Mpc on a side). The circle represents the actual size of the bubble around the green dot.}
    \label{fig:MetalGridZoom}
\end{figure*}

\begin{figure*}
    \centering
    \includegraphics[width=\textwidth]{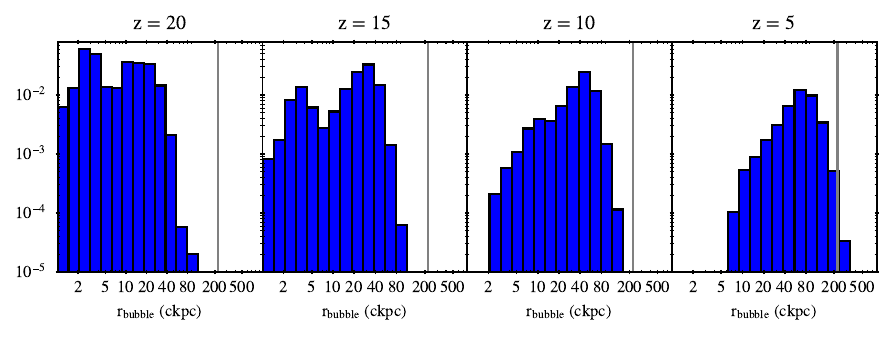}
    \caption{Distribution function of the metal bubble size computed according to Eq. \ref{eq:Rbubble} at 4 different redshifts. The grey column indicates the predicted value for the maximum bubble size at z = 6 from \citet{Trenti2009}. }
    \label{fig:BubbleDF}
\end{figure*}

\begin{figure*}
    \centering
    \includegraphics[width=\textwidth]{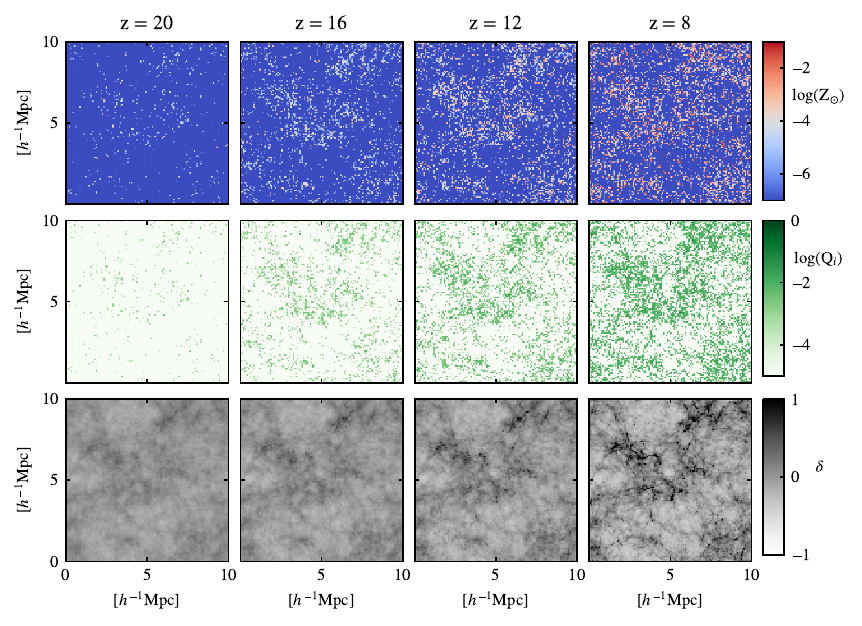}
    \caption{2D projections of the IGM metallicity $\log (Z_{\rm IGM})$ in solar units (top row), the metal filling factor $\log(Q)$ (middle row) and the overdensity $\delta$ (bottom row). From left to right, the columns correspond to the same redshift: $z = 20, 16, 12, 8$. Each slice is 10 $h^{-1}$ Mpc on a side and considers the average contribution over the entire thickness of the box. In the top row, red (blue) represents enriched (pristine) pixels: $Z_{\rm IGM} \geq (<) Z_{\rm crit}$. Green pixels have a larger filling factor $Q$, indicating a larger probability of a galaxy accreting enriched gas. Dark cells in the bottom row have larger mean overdensity $\delta$.}
    \label{fig:MetalMaps}
    %$\sim0.62 h^{-1}$Mpc thick.
\end{figure*}

%This procedure for external metal enrichment is significantly faster than computing the distance between all the pairs of galaxies while remaining statistically consistent (see Appendix \ref{sec:AppendixA} for a more detailed discussion).
We illustrate an example in Fig. \ref{fig:MetalGridZoom}, where we selected a 2D slice of our L10 box at z = 10 with a thickness of $0.078 h^{-1}$cMpc. The red (blue) dots are the galaxies with $M_{\rm vir} \geq 10^6 M_{\odot}$ (not) externally metal-enriched, while the green dots are previously formed galaxies with r$_{\rm bubble} \geq 0.05 h^{-1}$ cMpc. For a better visualization, we zoomed into a slice with $0.24 h^{-1}$ cMpc on a side showing the actual size of the metal bubble (right plot). Inside this region, all the 7 galaxies that formed inside the bubble are actually marked as externally enriched. At the same time, the 5 galaxies that are marked as not enriched lie outside the bubble, showing that the adopted approximation works quite well (see Appendix \ref{sec:AppendixA} for a more detailed discussion). This demonstrates that this method can reproduce the statistical properties of inhomogeneous metal enrichment. The main limitation of this technique is that we are not accounting for the overlap of the bubbles and thus are overestimating the metal-filling factor and underestimating the metallicity of those galaxies that are polluted from more than one galaxy. However, given the small size of the bubbles (see discussion in Section \ref{sec:global} and Fig. \ref{fig:BubbleDF}), this is not a major factor, especially prior to reionization. We note that the methodology shown here is similar to \citet{Sassano2021} with the main differences being that we divided the volume of our simulation in cells allowing a more precise computation of the metallicity of the enriched regions.\\
Fig. \ref{fig:BubbleDF} shows the bubble size distribution function (r$_{\rm bubble}$ in ckpc) at redshift 20, 15, 10 and 5. From this plot, we see the bubble growth over time from a typical radius of $\sim 30$ ckpc at $z = 20$ to $\sim 150$ ckpc at $z = 5$. These values are consistent with the results shown in \citet{Trenti2009} (r$_{\rm bubble} \leq 150 h^{-1}$ ckpc at $z = 6$) with only a few bubbles at $z = 5$ larger than this value. \\
Fig. \ref{fig:MetalMaps} shows the IGM metallicity (top row), metal filling factor (middle row) and density (bottom row) maps at four different snapshots. The red (blue) cells in the top row have $Z_{\rm IGM} \geq (<) Z_{\rm crit}$ indicating that galaxies that will be externally polluted in those cells will accrete enriched (pristine) gas and will form Pop. II (III) stars. Comparing the maps at the same snapshot, we see that the more overdense regions have a larger filling factor and a higher metallicity as these hosts more galaxies, hence more metal bubbles. Fig. \ref{fig:MetalMaps} also shows the progressive enrichment of the IGM over the cosmic history since, as we move to lower redshift (from left to right panels) we have more enriched cells and with larger filling factors. 
%Finally, we note that the metallicity map at $z = 5$ has less enriched pixels than the map at $z = 10$. This happens because when a galaxy becomes a satellite, all its ejected and hot reservoir (gas and metals) is associated to the hot gas of the central galaxy hence the IGM metallicity (that is computed from the metals ejected from all the galaxies) will decrease.

%This part is almost done. What you might want to add here are the resolution tests. Not sure though if 1) is worth to add those (probably yes) and 2) they should be here or in the following section.
%I don't know if it's worth to add the fitting formula for the 2 point CF. 
%Probably it's worth showing the bubble distribution function at different redshift here. If you do that and you want to discuss the resolution of the grid, it makes sense to do it just after BDF.

\subsection{Pop. III star formation}
\label{sec:PopIII}
%With all the physics described in the previous sections in place, it is straightforward to understand whether a galaxy in \textsc{meraxes} will form Pop. III or Pop. II stars. A galaxy that already formed stars or a galaxy that formed in a region enriched with metals so that its metallicity is larger than $Z_{\rm crit} = 10^{-4}Z_{\odot}$ will form Pop. II stars. We adopted the same prescription and parameters for Pop. II star formation as described in M16 and \citet{Qiu2019}. The uncertainty around the properties of Pop. III stars motivated us to include Pop. III star formation in a flexible way so that it is easy and fast to investigate the impact of Pop. III stars changing a few parameters.\\
In this work we modify Pop. II description in M16 to allow star formation in Pop. III galaxies (see Section 2.4 in M16). In particular, once a galaxy reaches a mass of cold gas larger than a critical value $m_{\rm crit}$, this galaxy will convert the cold gas into stars according to the star formation efficiency $\alpha_{\rm SF}$. The value of $m_{\rm crit}$ is computed as in M16 and is determined by a critical surface density $\Sigma_{\rm norm}$ of the gas in the cold disk. Since we are now considering two different stellar populations, we adopted two different free parameters, both for the star formation efficiency $\alpha_{\rm SF, III}$ and for the critical surface density $\Sigma_{\rm crit, III}$. Recent full hydrodynamical simulations that follow the collapse of a pristine gas cloud until a Pop. III star is formed \citep[e.g.][]{Hirano2014,Stacy2016,Chon2021}{}{}, suggest that the Pop. III IMF is shifted to larger masses as a result of less fragmentation due to inefficient cooling. Within this work, we adopted the IMFs from \citet{Raiter2010} (see Table \ref{tab:IMFparams}), while for Pop. II stars, we assumed a \citet{Kroupa2001} IMF. For the fiducial model, we chose a Salpeter IMF with a mass between 1 and 500 M$_{\odot}$. The choice of the IMF is crucial as it determines many properties of the stellar population, including what fraction of stars will explode as supernovae and, hence, the amount of energy and metals injected into the IGM. In this work, we explore the impact of the 4 free parameters discussed above (see also Table \ref{tab:PopIIIparams}) since these have the strongest impact on the Pop. III star formation history. \textsc{meraxes} includes more free parameters describing the energy coupling efficiency of the supernova explosion (see M16 and \citealt{Qiu2019}) that will not be explored in this work (we will take the same fiducial values as for Pop. II). Since the focus of this work is on the first galaxies formed during the Cosmic Dawn, we do not explore parameters describing the UV ionizing and X-ray radiation (e.g. escape fraction), and we will take the same fiducial values as in the previous works (see \citealt{Balu2023}). Those will be discussed in a future work focused on the EoR using a larger box. \\
%Within this work, Pop III star will form according to a simple power-law IMF: $\phi(M)dM = AM^{\alpha}dM$ with $M\in[M_{\rm min}, M_{\rm max}]$. We allow $\alpha$, $M_{\rm min}$ and M$_{\rm max}$ to be free parameters that can be changed between the different runs while A is the normalization constant determined from $\int_{M_{\rm min}}^{M_{\rm mas}}M\phi(M)dM = 1$. Summarizing, Pop III star formation is regulated by only 4 free parameters: $\alpha_{\rm SF, III}, \alpha,$ M$_{\rm min}$, M$_{\rm max}$.\\
The fate of a zero-metallicity star is quite uncertain due to the many poorly constrained physical mechanisms. Here, we adopt a simplified picture of the final fate of a Pop. III star depends only on its initial mass (\citealt{Heger2002}). For masses below 8 M$_{\odot}$, there will be no SN event. If M$_{\star} \in [8,40]$M$_{\odot}$ it will explode as a core-collapse SN (CCSN) leaving a remnant, if M$_{\star}\in [140,260]$M$_{\odot}$ there will be a pair-instability SN (PISN) leaving no remnant and if M$_{\star}\in [40,140]$M$_{\odot}$ or M$_{\star}$>260M$_{\odot}$ stars collapse directly into a black hole (BH) with negligible feedback.
Given the different endings of a Pop. III star life, the choice of the IMF is crucial to compute the timing of delayed supernova feedback. While we use the technique of \citet{Qiu2019} for Pop. II stars with precomputed tables assuming a Kroupa IMF (\citealt{Kroupa2001}), for Pop. III supernova feedback we estimated the amount of SN energy and metal yields with an analytic calculation as in M16. The total energy provided by Pop. III supernovae explosions at the snapshot j is:
\begin{equation}\begin{split}
     \Delta E_{\rm total,III,j} & = \sum_{i = j - 17}^{i=j} \bigg(\Delta E_{\rm CCSN,III,j} + \Delta E_{\rm PISN,III,j}\bigg) \\ 
    & =\epsilon_{\rm energy}\bigg(\sum_{i = j - 17}^{i=j}\bigg(\Delta m^{\rm i}_{\star,\rm III}\eta_{\rm III,CCSN,j}^{\rm i}E_{\rm CC}\bigg)\\
    & + \Delta m^{\rm j}_{\star,\rm III}\eta_{\rm III, PISN} E_{\rm PISN}\bigg),
    \label{eq:delayed_feed}
\end{split}\end{equation}
    %\Delta E_{\rm total,III,j} & = \Delta E_{\rm CCSN,III} + \Delta E_{\rm PISN,III} \\ 
    %& =\Delta m_{\star,\rm III}\epsilon_{\rm energy}(\eta_{\rm III,CCSN}E_{\rm CC} + \eta_{\rm III, PISN} E_{\rm PISN}),
%\end{split}\end{equation}
where $\Delta m_\star^{\rm i }$ is the new stellar mass formed at an earlier snapshot i, $\epsilon_{\rm energy}$ is the energy coupling efficiency (\citealt{Qiu2019}), $E_{\rm CC} = 10^{51}$ erg and $E_{\rm PISN} = 10^{53}$ erg are the typical energy for a core collapse and a pair-instability supernova respectively and 
\begin{eqnarray}
    \eta_{\rm III,CCSN,j}^{\rm i} = \int_{m_{\rm min,j}^i}^{m_{\rm max,j}^i}\phi(M)dM \\
    \eta_{\rm III,PISN} = \int_{140M_{\odot}}^{260M_{\odot}}\phi(M)dM
    \label{eq:mass_limit}
\end{eqnarray}
are respectively the fraction of stars formed during snapshot i that go core collapse and pair-instability supernova in snapshot j. These are computed by integrating the chosen IMF over the correct mass limits. Assuming that all-star formation occurred in the middle of the snapshot, ${m_{\rm min,j}^i}$ and ${m_{\rm max,j}^i}$ for CCSN are computed from the lifetimes for Pop. III stars using \citet{Schaerer2002} assuming no mass loss and zero metallicity. We highlight that a zero metallicity star with $M \in [140,260]$ has a lifetime smaller than the time separation between two consecutive snapshots in \textsc{meraxes}; hence, it will explode as a PISN in the same snapshot in which it forms. For this reason, in Eq. \ref{eq:mass_limit}, the mass limit for $\eta_{\rm PISN}$ is the entire mass range of the PISN and when computing the total energy from PISN injected at snapshot j in Eq. \ref{eq:delayed_feed} we consider only the stars that are formed at snapshot j. For CCSN, instead, we keep track of the star formation history over the last 17 snapshots, which correspond to $\gtrsim 40$ Myr, after which all stars with $M\geq 8M_{\odot}$ will already be exploded. To compute the amount of gas that is reheated and ejected from the halo, we adopted the same prescription as \citet{Qiu2019}.\\
We also update the amount of ejected gas and metals from Pop. III stars. These are taken from \citet{Heger2010} assuming non-mixing (no rotation) and a supernova explosion of 1.2 $\times 10^{51}$ erg. As in M16, when a star goes core-collapse supernova at snapshot j, it will release an amount $\Delta m_{\rm CCSN, Z,j}$ of metals and $\Delta m_{\rm CCSN, X,j}$ of gas into the interstellar medium:
\begin{equation}
   \Delta m_{\rm CCSN,Z,j} = \sum_{i = j - 17}^{i=j}\frac{\Delta m_{\rm CCSN, Z, j}^i}{m_{\rm CCSN}}\Delta m_\star,
   \label{eq:SNmetals}
\end{equation}
where $m_{\rm CCSN}$ is the total fraction of stars that ends as a CCSN:
\begin{equation}
    m_{\rm CCSN} = \int_{8M_{\odot}}^{40M_\odot}M\phi(M)dM
\end{equation}
and $\Delta m_{\rm CCSN, Z;i,j}$ is the amount of metals released by a star formed at snapshot $i$ and going supernova at snapshot j:
\begin{equation}
    \Delta m_{\rm CCSN, Z, j}^i = \int_{m_{\rm min,j}^i}^{m_{\rm max,j}^i}M\phi(M)ZdM.
    \label{eq:SNyields}
\end{equation}
In the above equation, Z is the metal mass fraction released into the ISM from supernovae, and it is incorporated inside the integral because it is not fixed over the mass range (i.e in stars with $M_\star \geq 25 M_{\odot}$ most of the metals will fall onto the remnant due to the massive fallback). With an entirely analogous calculation, it is possible to find $\Delta m_{\rm CCSN, X,j}$ (we just need to replace Z with the gas mass fraction X). Once again, we are doing delayed SN recycling only for CCSN because, in the case of pair instability, all the feedback is contemporaneous. For PISN, we considered the gas and metal yields from \citet{Heger2002}.\\
In all the previous \textsc{meraxes} works, all the stellar mass locked up in remnants (neutron stars and BHs) was neglected as it was recycled into the gas mass budget of the galaxy. We decided to drop this approximation for Pop III stars as they are likely to leave more massive remnants (some might be massive enough to be the first "light" seeds of supermassive BHs). Firstly, we need to consider the BHs that formed after a "failed SN scenario" typical of a star with an initial mass $M_\star \in [40,140]M_\odot$ and $M_\star > 260 M_\odot$. The BH mass arising from this scenario will be:
\begin{equation}
    m_{\rm BH} = \Delta m_\star\bigg(\int_{40M_{\odot}}^{140M_\odot}M\phi(M)dM + \int_{260M_\odot}^{M_{\rm max}}M\phi(M)dM\bigg).
\end{equation}
Finally, we need to account for the BH remnants that are left after a CCSN. In this case, we use equations \ref{eq:SNmetals} and \ref{eq:SNyields} substituting the metal mass fraction with the BH mass fraction $B = (1 - X - Z)$ (since all the remaining material that has not been ejected is locked onto the stellar remnant). \textsc{meraxes}, does not evolve the remnants (via accretion), as the main focus of this work is Pop. III stars. However, the impact of the first accreting BHs on the formation of the supermassive black holes and radiative backgrounds might be important even at high-redshift (\citealt{Ventura2023}).\\
%thus in a future work we are planning to follow their evolution.\\

\begin{table*}
    \centering
    \caption{Free parameters of Pop. III star formation.}
    \label{tab:PopIIIparams}
    \begin{tabular}{cccccccccc}
        \hline
        Parameter & Description & Fiducial value (Pop II) \\
        \hline
        $\alpha_{\rm SF}$ & Star formation efficiency & 0.008 (0.08) \\
        Z$_{\rm crit}$ & Critical metallicity for Pop III star formation & $10^{-4}$Z$_{\odot}$ \\
        $\Sigma_{\rm norm}$ & Critical surface density of cold gas for Pop III star formation & 0.37 (0.37) M$_\odot$ pc$^{-2}$ \\
        IMF type$^{\star}$ & Shape of the Initial Mass Function & Sal (Kroupa)\\
        \hline
        $^\star$see Table \ref{tab:IMFparams}
    \end{tabular}
\end{table*}

\begin{table}
    \centering
    \caption{IMF model parameters.}
    \label{tab:IMFparams}
    \begin{tabular}{cccccccccc}
        \hline
        Label & Type & M$_{\rm low}$ & M$_{\rm up}$ & $\alpha$ & M$_{\rm c}$ & $\sigma$ \\
        \hline
        Sal & Salpeter & 1 & 500 & -2.35 \\
        logA & log-Normal & 1 & 500 & & 10 & 1.0 \\
        logE & log-Normal & 1 & 500 & & 60 & 1.0 \\
        \hline
    \end{tabular}
\end{table}

\subsection{Pop. III luminosity}
\label{sec:PopIII_spectra}
\begin{figure}
    %\centering
    \includegraphics[width=\columnwidth]{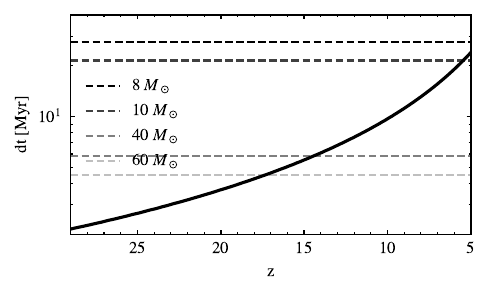}
    \caption{Duration of the snapshot dt in Myr as a function of redshift. Dashed lines correspond to the lifetime of a zero metallicity star of 8M$_\odot$, 10M$_\odot$, 40M$_\odot$ and 60M$_\odot$.}
    \label{fig:cadence}
\end{figure}

In order to investigate the observability of the first Pop. III galaxies we implement spectral energy distributions (SED) for Pop. III stars. We use SEDs from \citet{Raiter2010} 
%(based on the Yggdrasil model \citealt{Zakcrisson2011}) 
that have been computed for the IMFs listed in Table \ref{tab:IMFparams} assuming that star formation occurs in an instantaneous burst (see \textit{instantaneous burst} in \citealt{Raiter2010}). These SEDs also include the nebular continuum emission and the UV ionizing properties. We used the latter to compute the number of UV ionizing photons per baryon $n^{\gamma}_b$ for each IMF provided. To compute the luminosity of galaxies at a specific wavelength $\lambda$, we used the model introduced by \citet{Qiu2019} (see also \citealt{Mutch2023}), and we extended it to Pop. III galaxies. Having the Pop. III SEDs, we can compute the luminosity of a Pop. III galaxy at time t as:

\begin{equation}
    L_\lambda(t) = \int_0^t\psi(t-\tau)S_\lambda(\tau) T_\lambda(\tau)d\tau.
    \label{eq:PopIIILum}
\end{equation}
Here $\tau$ is the stellar age, $\psi(t - \tau)$ the stellar mass formed at $t - \tau$ with age between $\tau$ and $\tau + d\tau$, $S_\lambda(\tau)$ is the luminosity of a stellar population per unit mass and $T_\lambda(\tau)$ is the transmission function of the interstellar medium. For the latter term, we refer the reader to \citet{Qiu2019} while $S_\lambda$ for each Pop. III IMF has been taken from \citet{Raiter2010} as discussed above. This calculation is nearly identical to \citet{Qiu2019}, with the difference being that we do not have the metallicity dependence because Pop. III stars SEDs are defined with a zero metallicity. In this framework, we assume that the star formation occurs continuously throughout the snapshot. To compute $\psi(t - \tau)$, we take the stellar mass formed in that snapshot and average it over the entire duration of the snapshot. While this is a good approximation for Pop. II stars, Pop. III star formation is expected to occur in a single burst because of the feedback from Pop. III stars are likely to prevent continuous star formation. For this reason, we expanded the calculation of the luminosity of Pop. III galaxies assuming that new stars $\Delta m_\star$ form instantaneously at the time $t - \tau = t_0$. Hence, we can write $\psi(t - \tau) = \Delta m_\star \delta(t_0)$ and Eq. \ref{eq:PopIIILum} simplifies to:

\begin{equation}\begin{split}
    L_\lambda(t) &= \Delta m_\star\int_0^td\tau\delta(t - t_0)S_\lambda(\tau)T_\lambda(\tau) \\
    & = \Delta m_\star S_\lambda(t - t_0)T_\lambda(t - t_0).
    \label{eq:PopIIIinst}
\end{split}\end{equation}
As we will discuss in Section \ref{sec:observability}, instantaneous (instead of continuous) Pop. III star formation has an impact on the estimated luminosity function of galaxies hosting Pop. III stars. This is because Pop. III stars have lifetimes that can be shorter than the duration of the snapshot (at $z = 10$ the snapshot will last 10 Myr, see Fig. \ref{fig:cadence}). Hence, when we consider continuous star formation, we average the star formation over the entire snapshot, leading to lower luminosity with a higher duty cycle. \\
%and we will lose the contribution coming from Pop. III stars that are dead by the end of the snapshot. \\

%When we instead consider Pop. III stars to form in a single burst, if this burst will occur toward the end of the snapshot the galaxy will appear much brighter when it is observed in that snapshot. \\ 
In order to account for instantaneous star formation, we assumed that a Pop. III star formation episode in a galaxy can occur at random $\Delta t \in [0, t_{\rm snap}]$ prior to the end of the snapshot. 
%As so, compared to the continuous star formation model, when a galaxy have a small $\Delta t$, depending on the IMF, it can be up to 3 magnitudes brighter (these results will be discussed in Section \ref{sec:observability}). 
For the detailed evaluation of the UV luminosity function accounting for the stochasticity in the time at which the burst of star formation occurs in different galaxies, we refer the reader to Appendix \ref{sec:AppendixB}.

\section{Global evolution of Pop. III stars}
\label{sec:global}
\begin{figure}
    %\centering
    \includegraphics[width=\columnwidth]{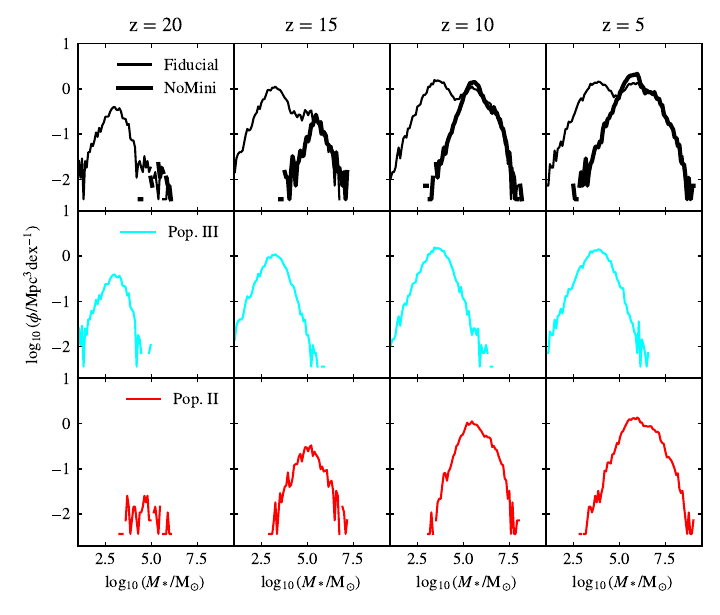}
    \caption{SMF at z = 20, 15, 10 and 5 for all stars (upper row), Pop. III stars (middle row) and Pop. II (lower row) using the fiducial model (Table \ref{tab:PopIIIparams}) and the NoMini model (thick line).}
    \label{fig:SMF_Fid}
\end{figure}

\begin{figure*}
    %\centering
    \includegraphics[width=\textwidth]{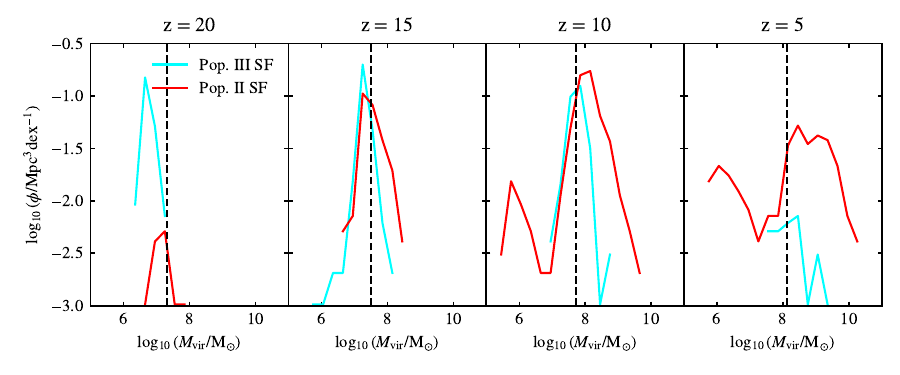}
    \caption{Halo mass function at z = 20, 15, 10 and 5 of Pop. III and Pop. II hosting systems. The dashed line corresponds to the $T_{\rm vir} = 10^4$ K that defines the atomic cooling limit.}
    \label{fig:HMF_Fid}
\end{figure*}

To explore the Pop. III contributions to the cosmic star formation history, we ran two simulations on the L10 box, one with all the updates described in Section \ref{sec:Model} adopting the fiducial parameters (see Table \ref{tab:PopIIIparams} and one without the new physics that we labelled as "NoMini"). Given that there are no observational constraints on Pop. III, the choice of a fiducial model is arbitrary. The one adopted in this work has both a low star formation efficiency and a Salpeter-like IMF, which will result in a relatively small global impact of Pop. III star formation compared to Pop. II. \\
In Figure \ref{fig:SMF_Fid} we show the stellar mass function (SMF) at different redshifts accounting for all (black), only Pop. III dominated (cyan), and only Pop. II dominated (red) galaxies. We classified a galaxy as Pop. III or Pop. II dominated based on which population is brighter in the UV band. Thin (thick) lines are computed from the fiducial (NoMini) model. 
The total SMF has two components, separated by their stellar types. In this fiducial model, as Pop. III SFE is 10 times lower than Pop. II, we see a resulting bimodality (which we will further expand discussion in Sec 3.2.1). 
As shown in the upper panels, the new updates on \textsc{meraxes} mostly affect the lower end of the SMF with a larger impact at high-$z$. This reflects the star formation in mini-halos (we are now considering molecular cooling), which is dominant at $z \geq 15$ before the Lyman-Werner background becomes strong enough to photodissociate all the molecular hydrogen. \\
%An interesting feature of Fig. \ref{fig:SMF_Fid} is that the total SMF now shows a bimodality with two distinct peaks separated by $\sim 2 $ dex. As shown in the middle panels, the low-mass peak of the SMF is dominated by Pop. III systems (at $z = 10$, the peak is centred at $M_\star \sim 10^{3-4}$M$_\odot$) and the more massive one by Pop. II. This bimodality of the SMF could have an impact on the faint end of the luminosity function at $z \sim 10$ (although the mass of these Pop. III systems is very small). We discuss the possible observability of these Pop. III dominated galaxies in section \ref{sec:observability}. 
The low mass of Pop. III systems is mainly a consequence of both the shorter lifetimes of Pop. III stars (given their larger mass) and the lower SF efficiency, but also suggests that most of the Pop. III star formation must occur in mini-halos. In order to investigate this, we checked which halos form Pop. III and Pop. II stars. In Fig. \ref{fig:HMF_Fid}, we show the halo mass function for Pop. III (cyan line) and Pop. II (red line) star-forming halos at $z = 20, 15, 10$ and 5. The dashed line corresponds to the limit of atomic cooling ($T_{\rm vir} = 10^4$K). Given the small size of the box, we have only a few atomic cooling halos at $z \sim 20$, so all star-forming systems are mini-halos (and are mostly Pop. III). At lower redshift, the peak of the distribution shifts to higher masses (by $z = 10$ most of the stars form in atomic cooling halos except for some low mass halos that are undergoing a merger event) and the impact of Pop. II increases. Finally, Fig. \ref{fig:HMF_Fid} shows that Pop. III stars mostly form in mini-halos (except at $z = 5$ when the radiative feedback has suppressed both the gas infall and cooling in mini-halos). This is because the more massive halos are more likely to have already experienced star formation and so will be internally chemically enriched. \\

\begin{figure}
    %\centering
    \includegraphics[width=\columnwidth]{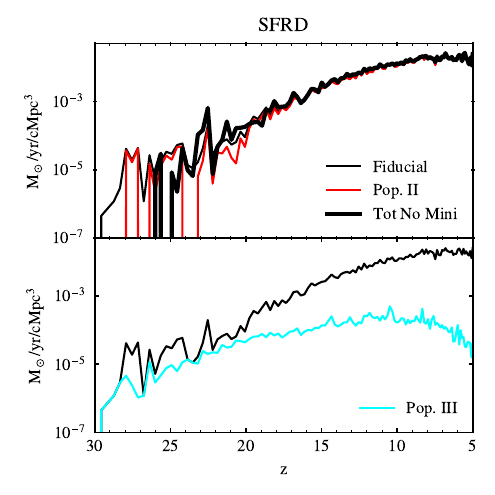}
    \caption{SFRD vs redshift for all stars (black line), Pop. III (cyan) and Pop. II (red) using the fiducial model (thin line) and the NoMini model (thick line).}
    \label{fig:SFRD_Fid}
\end{figure}

In Figure \ref{fig:SFRD_Fid} we show the total star formation rate density (SFRD) as a function of redshift for both the fiducial and the NoMini model. In the case of the fiducial model, we also show the Pop. III (Pop. II) contribution in the upper (lower) panel. Mini-halos have an appreciable impact on the SFRD only up to $z \geq 20$, while at lower redshift, most star formation occurs in atomic cooling halos (at $z < 15$, the total SFRD in the fiducial and NoMini model are superimposed). We see that accounting for Pop. III star formation in mini-halos is crucial during the Cosmic Dawn because Pop. III stars dominate the global star formation history at $z \geq 20$ and are still relevant up to $z = 18$. As we can see in the lower panel, the Pop. III SFRD flattens at $z \sim 18$ and starts to decrease at $z \sim 10$. The early flattening occurs due to the build-up of the Lyman-Werner background that affects Pop. III star formation in minihalos. The sharp drop at $z \leq 10$ is mostly caused by the photoionizing feedback from reionization that also affects the atomic cooling halos. The feedback from reionization also mildly affects Pop. II as can be seen from the flattening of the red line at $z \sim 8$ in the upper panel\footnote{Given the small size of the box, the feedback from reionization it is slightly overestimated as we \textit{(i)} are missing the most massive systems that would not be affected by the photo-ionizing feedback and \textit{(ii)} the box gets completely ionized at $z \sim 6.9$ which is much earlier than what suggested by the latest Lyman-$\alpha$ forest measurements \citep[e.g.][]{Qin2021, Bosman2022}.} \\ %6.91

\subsection{Chemical enrichment of the IGM}

Compared to reionization, the chemical enrichment of the IGM is a much slower process due to the lower velocity of the expansion of the metal bubbles. As already found in previous works \citep[e.g.][]{Visbal2020,Yamaguchi2023}, the impact of external metal enrichment is subdominant compared to internal metal enrichment. However, external enrichment might still be important for low-mass halos that do not form stars until $z \sim 10$. 

\begin{figure}
    %\centering
    \includegraphics[width=\columnwidth]{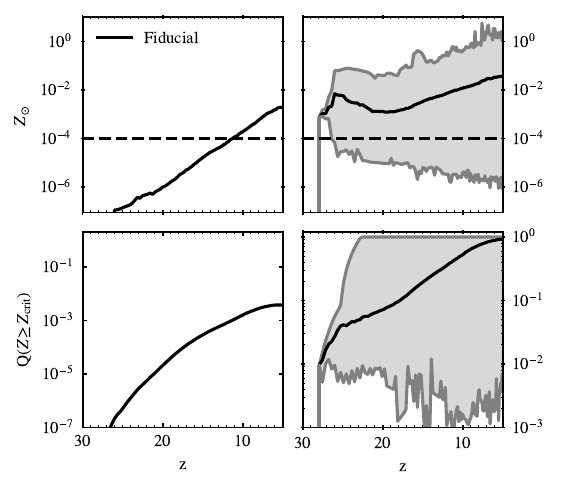}
    \caption{{\textbf{top:}} Average metallicity of the IGM (in solar units) vs redshift highlighting $Z_{\rm crit}$ for the Pop. III/II transition. {\textbf{bottom:}} fraction of the volume with a metallicity $Z \geq Z_{\rm crit}$ vs redshift. The left panels show these quantities averaged through the entire box, while the right panels consider only those cells with at least one metal bubble in them. Shaded regions in the right panels highlight the minimum and the maximum value (see text).}
    \label{fig:IGM_Fid}
\end{figure}

The top left panel of Fig. \ref{fig:IGM_Fid} shows the redshift evolution of the average metallicity of the box (in solar units) computed using Eq. \ref{eq:ZIGM} and averaging through all $128^3$ pixels. As expected, the average metallicity increases monotonically. It crosses the critical value at $z \sim 11$ and at the end of the simulation is $\sim 5 \times 10^{-3} Z_{\odot}$. The redshift evolution of $Z_{\rm IGM}$ is in very good agreement with \citet{Yamaguchi2023}, especially for their "bursty" models, which is the one that most closely resembles the star formation in \textsc{meraxes}. 
%However this quantity might be misleading as it seems to suggest that Pop. III stars are likely to be dominant down to $z \sim 11$ while we saw from Fig. \ref{fig:SFRD_Fid} that they become subdominant at a much higher-$z$. 
However, the average metallicity of the IGM does not completely inform us of the average metallicity of the galaxies since, when averaging through the entire box, we are considering voids that have no galaxies and hence zero IGM metallicity.
%This happens because when averaging through the entire box we are considering voids (underdense regions with no galaxies) that have zero metallicity. 
For this reason, in the top right panel, we computed the black solid line, averaging only through cells that have at least one metal bubble (meaning that there must be at least one galaxy that formed stars). The shaded region shows the scatter in the metallicity of these cells from the highest to the lowest metallicity. The average metallicity, in this case, is much higher, and it is always above the critical value. This suggests that once the first galaxies form, the ejection of metals from Pop. III is quite effective in the nearby regions, allowing a fast Pop. III/II transition in those cells where star formation already occurred (and the galaxies are internally enriched). 
%Moreover, the black line does not monotonically increase until $z \sim 20$ reflecting the large scatter in the metallicity between different cells (down to $z \sim 20$ each cell has only one or very few galaxies forming stars). 
The lower panels instead show the fraction of the IGM that reached the critical metallicity (or equivalently, the metal filling factor $Q(Z \geq Z_{\rm crit})$). Considering the entire box (left panel), less than 1\% of the volume gets enriched above $Z_{\rm crit}$ by $z \sim 5$. This result is fairly consistent with \citet{Visbal2020} and \citet{Yamaguchi2023} ($\sim 1\%$ by $z = 6$). These small differences are likely due to differences in the star formation models and in the choice of parameters, such as the star formation efficiency. If we focus only on the regions with at least one galaxy that formed stars, we find larger filling factors, and by the end of the simulation, those pixels are all completely enriched. This final result reflects our choice of pixel size that is designed to be similar to the average volume of the metal bubble at $z \sim 5$.

\begin{figure*}
    \includegraphics[width=\textwidth]{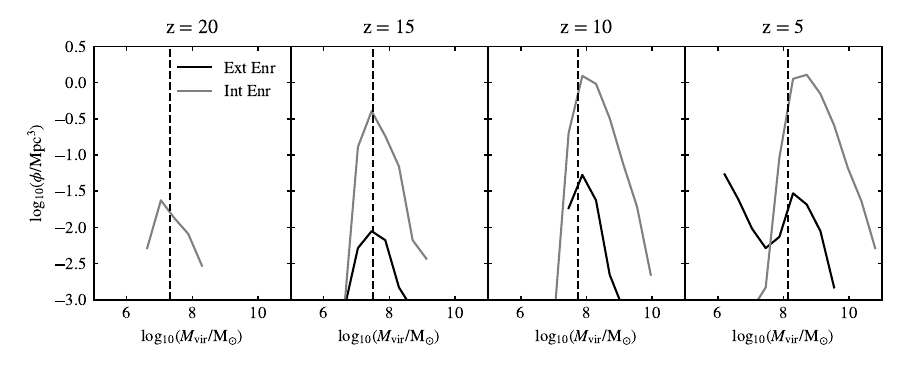}
    \caption{Halo mass function at z = 20, 15, 10 and 5 of Pop external (black) and internal (grey) metal enriched halos. The dashed line corresponds to the $T_{\rm vir} = 10^4$ K that defines the atomic cooling limit.}
    \label{fig:HMF_EnrFid}
\end{figure*}

\begin{figure}
    \includegraphics[width=\columnwidth]{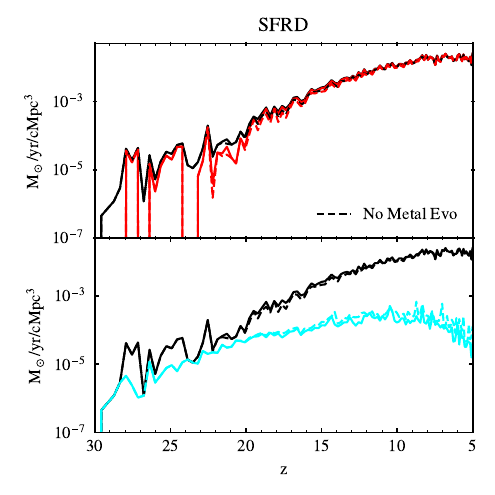}
    \caption{As Fig. \ref{fig:SFRD_Fid} for the fiducial model (solid lines) and turning off the external metal enrichment (dashed lines).}
    \label{fig:SFRD_NV}
\end{figure}

Finally, in Fig. \ref{fig:HMF_EnrFid}, we show the halo mass function for the externally (black line) and internally (grey line) metal enriched halos at $z = 20, 15, 10$ and 5\footnote{While physically a halo could be both internally and externally enriched, here we are simplifying assuming that a halo is either internally or externally enriched.}. While most halos get internally enriched by their own star formation, at $z = 5$ halos with virial masses $M_{\rm vir} \leq 10^{7.5} M_{\odot}$ get their metals mostly from a nearby supernova bubble. This picture reflects the fact that low-mass halos during reionization are not able to form stars because of the LW and photo-ionizing feedback. Hence, those objects can be chemically enriched only from an external source.\\
In conclusion, when looking at the global evolution of star formation, the effect of the external metal enrichment is quite negligible as it is important only for low-mass halos at low redshift that will hardly form stars due to the radiative feedback effects. We verified this by running a simulation without external metal enrichment (a halo can only get enriched by its own star formation). We plot in Fig. \ref{fig:SFRD_NV} the total, Pop. III and Pop. II SFRD (dashed lines) compared to the fiducial model (solid lines). There are no appreciable differences between the two models except at $z \leq 8$ when the dashed line is slightly larger (about 0.1 dex difference). 

\subsection{Pop. III star formation parameters}
%The previous results showed us that Pop. III stars mostly form in mini-halos hence they dominate the SFRD evolution at high-$z$ and the low-mass end of the stellar mass function.
The results in previous sections were obtained adopting quite conservative assumptions for Pop. III stars given the very low SF efficiency and the Salpeter-like IMF (although with $M_{\rm min} = 1 M_\odot$). In the following sections, we explore the four main free parameters that regulate Pop. III star formation in \textsc{meraxes} listed in Table \ref{tab:PopIIIparams}: the Pop. III star formation efficient $\alpha_{\rm SF, III}$, the critical metallicity for Pop.III/II transition $Z_{\rm crit}$, the critical surface density of cold gas for Pop. III star formation to occur $\Sigma_{\rm norm, III}$ and the IMF.

\subsubsection{Pop. III SF efficiency}
The star formation efficiency determines the conversion of the cold gas into stars and is the free parameter with the largest impact on the Pop. III global SFRD and SMF. 
This parameter is largely unconstrained, with some simulations supporting very low values ($10^{-4} - 10^{-3}$ \citealt{Skinner2020}) and others suggesting higher values (\citealt{Fukushima2020}). 
%ADD HERE ABOUT VALUES OF SFEFF! Chon vs Skinner & Wise
We ran two simulations keeping all the Pop. III free parameters unchanged except for $\alpha_{\rm SF, III}$, which we boosted by one order of magnitude (to the same value we use for Pop. II $\alpha_{\rm SF} = 0.08$) and to a very high value ($\alpha_{\rm SF} = 0.5$).\\

\begin{figure}
    %\centering
    \includegraphics[width=\columnwidth]{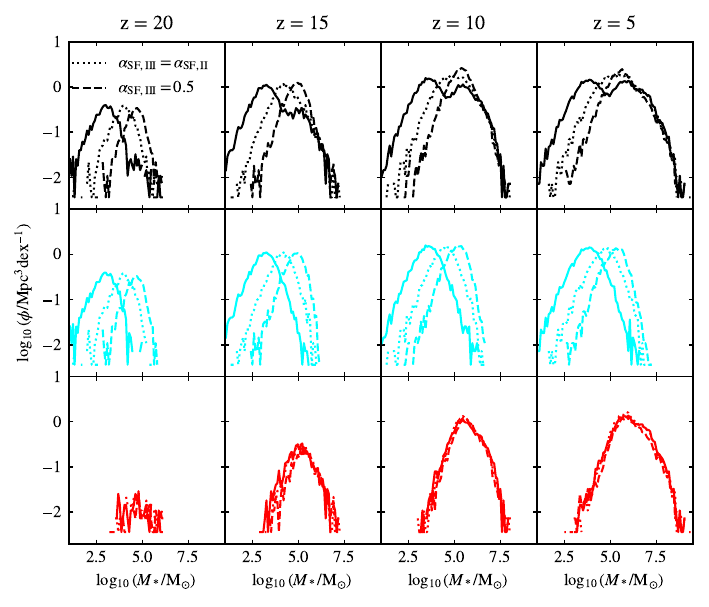}
    \caption{As Fig. \ref{fig:SMF_Fid} using the fiducial model (solid), $\alpha_{\rm SF, III} = \alpha_{\rm SF, II}$ (dotted) and $\alpha_{\rm SF, III} = 0.5$ (dashed).}
    \label{fig:SMF_SfEff}
\end{figure}

As shown in Fig. \ref{fig:SMF_SfEff}, Pop. III galaxies become more massive as $\alpha_{\rm SF}$ increases, erasing the "double peak" feature in the total SMF. For the model with $\alpha_{\rm SF, III} = \alpha_{\rm SF, II}$, the peak of the Pop. III SMF is still shifted to the left by half dex compared to the Pop. II. This is because, despite having the same star formation efficiency, Pop. III galaxies mostly form in mini-halos. For $\alpha_{\rm SF, III} = 0.5$, the peak of the Pop. III and Pop. II SMF is at the same mass value ($10^5 - 10^6 M_\odot$ at $z \sim 15 - 5$) with the only difference being the larger variance for Pop. II galaxies. Increasing the Pop. III SF efficiency does not affect the Pop. II SMF (except for small variations at $z = 5$) meaning that the mechanical feedback does not change significantly.\\

\begin{figure}
    \includegraphics[width=\columnwidth]{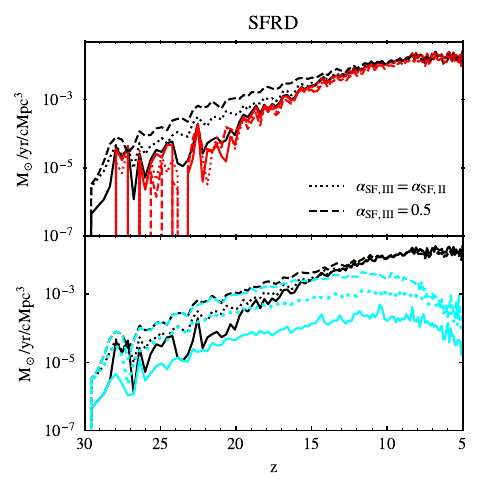}
    \caption{As Fig. \ref{fig:SFRD_Fid}  using the fiducial model (solid), $\alpha_{\rm SF, III} = \alpha_{\rm SF, II}$ (dotted) and $\alpha_{\rm SF, III} = 0.5$ (dashed).}
    \label{fig:SFRD_SfEff}
\end{figure}

The star formation efficiency also regulates the Pop. III/II transition as it directly correlates with the SFRD. Looking at the SFRD in the three different models (Fig. \ref{fig:SFRD_SfEff}) we see that for $\alpha_{\rm SF, III} = 0.5$, the SFRD is dominated by Pop. III up to $z \sim 15$ compared to the fiducial model where Pop. II SF becomes larger than Pop. III at $z > 20$. The Pop. II SFRD does not significantly change between the three models, and so the change in the Pop. III SFRD also affects the total SFRD (the dashed line is almost one order of magnitude higher than the solid one). All the models converge at $z \sim 13$ when even with the highest $\alpha_{\rm SF, III}$ the total SFRD is entirely dominated by the Pop. II contributions. 

\subsubsection{Critical metallicity}
The critical metallicity defines the metallicity at which there is a change in the IMF (metal-free gas clouds have inefficient cooling that leads to less fragmentation and higher masses \citealt{Chiaki2022}). There is currently no consensus on the value of $Z_{\rm crit}$, with two competing models that assume the fragmentation driven by either the carbon and oxygen line \citep[e.g.][]{Chon2021} or the dust cooling \citep[e.g.][]{Schneider2012,Chiaki2022}. The first class of models determine a $Z_{\rm crit} \sim 10^{-2} - 10^{-3} Z_\odot$ while the seconds give a lower value ($\sim 10^{-6} Z_\odot$). Some studies also argue that the value of $Z_{\rm crit}$ evolves with redshift due to the effect of the CMB (\citealt{Chon2022}). Following simulations in the literature (\citealt{Schneider2006, Visbal2020}), in this work we adopted an intermediate value $Z_{\rm crit} = 10^{-4} Z_\odot$ as the fiducial value, hereafter we will also show results for the two extreme values of $10^{-2} Z_{\odot}$ and $10^{-6} Z_{\odot}$.  

\begin{figure}
    %\centering
    \includegraphics[width=\columnwidth]{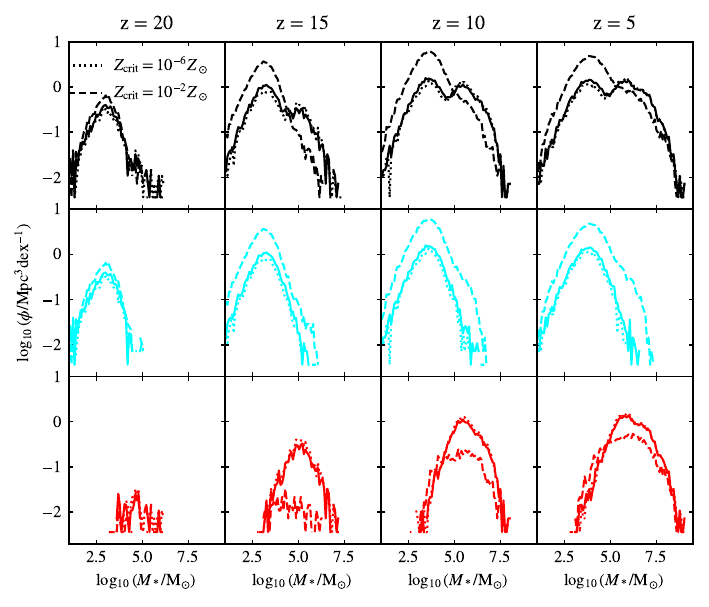}
    \caption{As Fig. \ref{fig:SMF_Fid} using the fiducial model (solid), $Z_{\rm crit} = 10^{-6}Z_\odot$ (dotted) and $10^{-2}Z_\odot$ (dashed).}
    \label{fig:SMF_Zcrit}
\end{figure}

The SMF between the fiducial model and Z$_{\rm crit} = 10^{-6}$ Z$_\odot$ does not change. This is because, as we saw in the top right panel of Fig. \ref{fig:IGM_Fid}, in the overdense regions where there is star formation, the metallicity becomes larger than $10^{-4} Z_\odot$ very quickly so that changing $Z_{\rm crit}$ to $10^{-6} Z_\odot$ does not further accelerate the Pop. III/II transition. However, considering a higher value changes the total SMF as we end up with many more Pop. III stars. As shown in Fig. \ref{fig:SMF_Zcrit}, in the Pop. III SMF in this case, many of the previously identified as Pop. II galaxies are now Pop. III galaxies across the entire mass range. The high-mass tail of Pop. III SMF extends up to $10^{7.5} M_\odot$ at $z = 5$ when $Z_{\rm crit} = 10^{-2} Z_\odot$, while the Pop. II systems are less massive as they start to form later and so have less time to build up their mass.  
The combined effect of Pop. III and Pop. II makes the total SMF computed from the simulation with $Z_{\rm crit} = 10^{-2} Z_\odot$ single peaked as the high-mass peak coming from Pop. II stars are washed out even at $z = 5$ so that the total SMF peaks at $M_\star \sim 10^4M_\odot$ at $z = 10 - 5$ with a high-mass tail that extends up to $10^9 M_\odot$. 
 
\begin{figure}
    \includegraphics[width=\columnwidth]{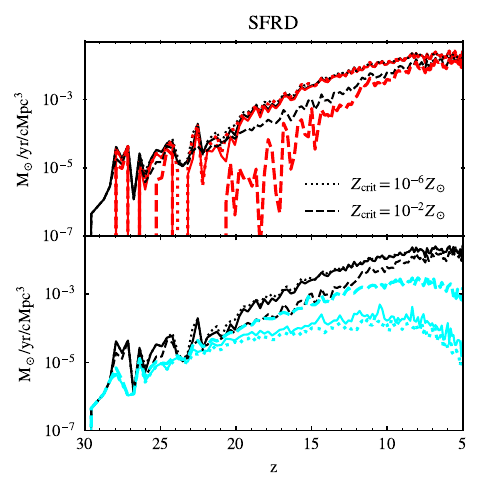}
    \caption{As Fig. \ref{fig:SFRD_Fid}  using the fiducial model (solid), $Z_{\rm crit} = 10^{-6}Z_\odot$ (dotted) and $10^{-2}Z_\odot$ (dashed).}
    \label{fig:SFRD_Zcrit}
\end{figure}

The critical metallicity also has a strong impact on the evolution of the SFRD (see Fig. \ref{fig:SFRD_Zcrit}) as it affects both the internal and external enrichment. While decreasing the value of Z$_{\rm crit}$ from the fiducial value only mildly decreases the Pop. III SFRD without altering the total result, choosing Z$_{\rm crit} = 10^{-2}Z_\odot$ strongly changes the SFRD history at $z \leq 18$. The Pop. III SFRD increases by 1-2 orders of magnitude while the Pop. II decreases by the same amount. It also starts to increase steadily only from $z \sim 18$. 
%These results indicate that both the external and the internal enrichment are affected by the increased critical threshold. 
Increasing the Pop. III star formation and simultaneously decreasing the Pop. II impacts on the total SFRD. Since Pop. III stars form less efficiently ($\alpha_{\rm SF, III} = 0.1 \times \alpha_{\rm SF, II}$) we predict a lower total SFRD compared to the fiducial model (and in this case, the two models do not converge until the very end of the simulation). Overall the critical metallicity heavily impacts both the SFRD and SMF during the Cosmic Dawn and the Epoch of Reionization and affects when the universe transitions from being Pop. III dominated Pop. II.
 
\subsubsection{Critical surface density}
\begin{figure}
    \includegraphics[width=\columnwidth]{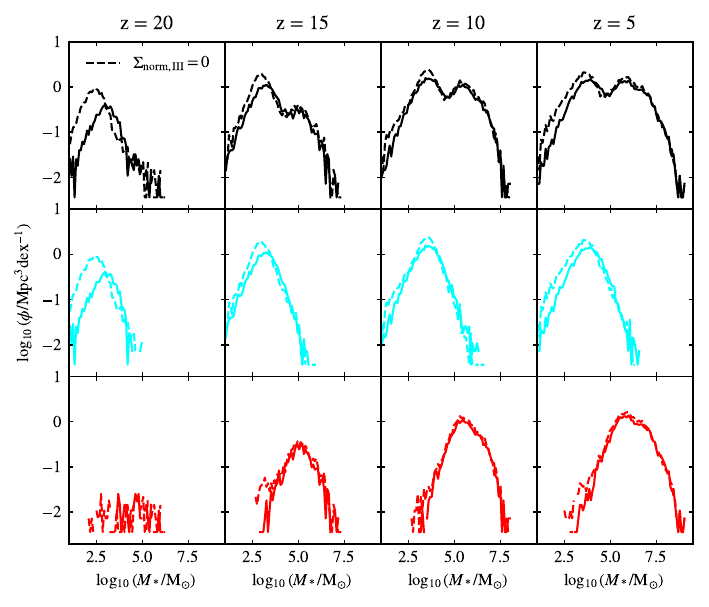}
    \caption{As Fig. \ref{fig:SMF_Fid} using the fiducial model (solid) and $\Sigma_{\rm norm, III} = 0$ (dashed).}
    \label{fig:SMF_SD0}
\end{figure}

In order to trigger the star formation in a galaxy, our model requires the gas density in the disk to be above a certain threshold. This result has been obtained in the observational results of \citet{Kennicutt1998, Kennicutt2021} and theoretically motivated by \citet{Kauffmann1996}; however, the value of this parameter is highly uncertain, especially at high-$z$ as it depends on the structure of the disk (e.g. thickness, turbulence etc.). We decided to duplicate this free parameter so that we have one for Pop. III and one for Pop. II. While keeping the Pop. II parameter fixed, we explore the extreme case of $\Sigma_{\rm norm, III} = 0$, which is equivalent to assuming that Pop. III stars start to form as soon as the cooling of the gas begins, and all cold gas is available to form Pop. III stars (in proportion to the star formation efficiency). This choice results in a larger abundance of low-mass Pop. III systems at high-$z$ (the largest difference between the cyan dashed and solid line in Fig. \ref{fig:SMF_SD0} are at $z = 20$ and 15).
%as even systems with a low mass of cold gas will form stars and a low mass of gas will result in a low stellar mass. 

\begin{figure}
    \includegraphics[width=\columnwidth]{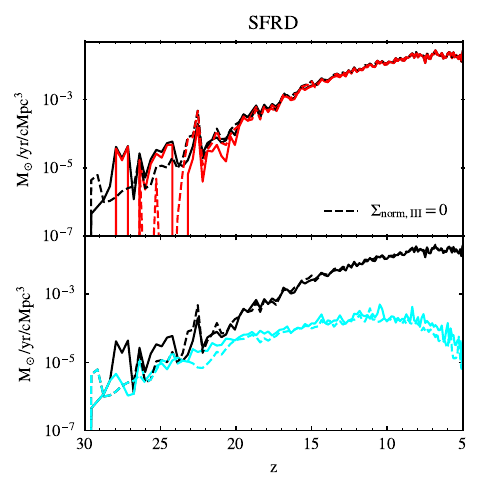}
    \caption{As Fig. \ref{fig:SFRD_Fid}  using the fiducial model (solid) and $\Sigma_{\rm norm, III} = 0$ (dashed).}
    \label{fig:SFRD_SD0}
\end{figure}

However, since the main changes in the Pop. III SF are in low-mass halos, setting $\Sigma_{\rm norm, III} = 0$ does not significantly change the SFRD history (see Fig. \ref{fig:SFRD_SD0} where the cyan dashed line is only slightly lower than the fiducial model).

Overall, this free parameter is the one with the smallest impact on the star formation history, as it only impacts the low-end of the stellar mass function at high-$z$ for which there are no observational constraints. 

%You might add a little discussion here about the first episodes of star formation.

\subsubsection{IMF}
The last parameter we explore is the IMF of Pop. III stars. Our fiducial model is quite similar to the Pop. II IMF as we assume a Salpeter IMF that favours the formation of low-mass stars. The only difference is that we take larger lower and upper limits (1 M$_\odot$ and 500 M$_\odot$). We consider two log-normal IMFs labelled logA and logE as in \citet{Tumlinson2006} with the parameters summarized in Table \ref{tab:IMFparams}. For these IMFs we have spectral energy distributions (from \citealt{Raiter2010}) which will be used in Section \ref{sec:observability}). The IMF logA is centred at $M = 10 M_\odot$, which enhances the probability of a Pop. III stars ending its life as a core-collapse SN. logE is centred at even a higher mass ($M = 60 M_\odot$), so we expect most of the stars with very short lifetimes to end their lives by collapsing into a black hole without any supernova explosion. Compared to the Salpeter IMF, both log-normal IMFs do not have many low mass stars ($M < 8 M_{\odot})$.

\begin{figure}
    \includegraphics[width=\columnwidth]{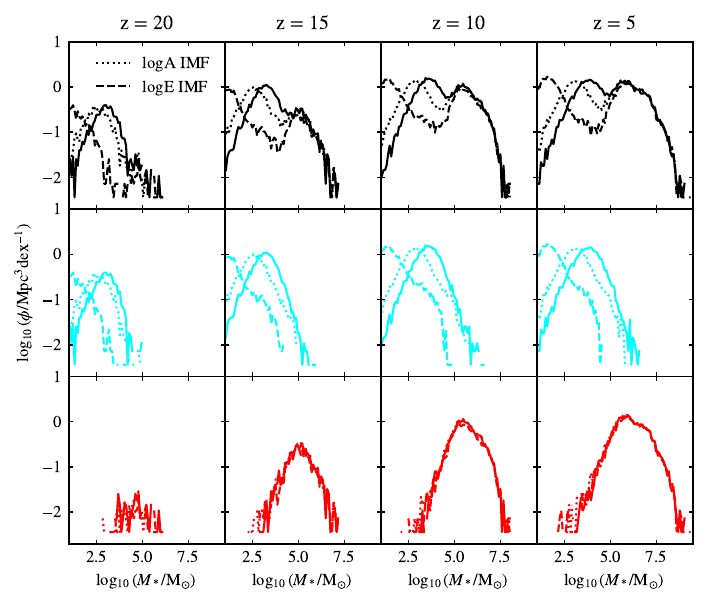}
    \caption{As Fig. \ref{fig:SMF_Fid} using the fiducial model (solid), the logA IMF (dotted) and the logE IMF (dashed).}
    \label{fig:SMF_IMF}
\end{figure}

Decreasing the lifetime of Pop. III stars make the Pop. III SMF in Fig. \ref{fig:SMF_IMF} shift to lower masses, increasing the separation between the two peaks of the total SMF (the effect of changing the IMF towards higher masses is the opposite of increasing the SF efficiency). The difference between the fiducial model and one adopting the logA IMF is less than one dex. However, in the logE IMF model, most of the stars die in the same snapshot in which they form (the lifetime of a $60 M_\odot$ star is $\sim 4.5$ Myr, which at $z < 15$ is less than the time separation between two consecutive snapshots in \textsc{meraxes}, see also Fig. \ref{fig:cadence}). This effect strongly prevents the build-up of Pop. III systems, and in Fig. \ref{fig:SMF_IMF}, we can see that the most massive Pop. III galaxies at $z = 5$ have $M \sim 10^5 M_\odot$. There are no differences in the Pop. II SMF with different IMFs.

\begin{figure}
    \includegraphics[width=\columnwidth]{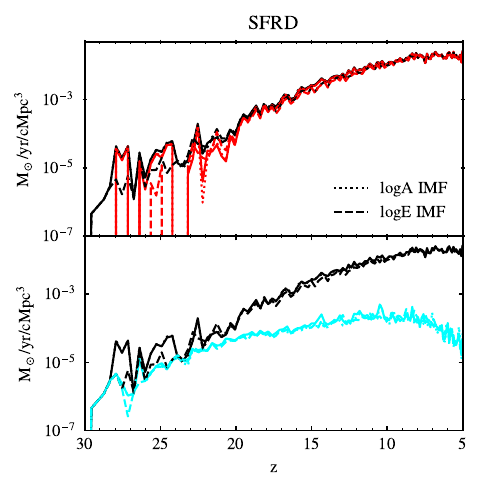}
    \caption{As Fig. \ref{fig:SFRD_Fid} using the fiducial model (solid), the logA IMF (dotted) and the logE IMF (dashed).}
    \label{fig:SFRD_IMF}
\end{figure}

In difference from the SF efficiency, the IMF mostly affects the stellar mass function. This is because the main change is with respect to the lifetime of Pop. III stars. In Fig. \ref{fig:SFRD_IMF} we see that there are no significant differences in the SFRD from adopting different IMFs, except for the first few snapshots of our simulation where the star formation is mildly reduced in the logE model. This result comes from the stronger feedback typical of the more top-heavy IMF (a larger number of massive stars implies more supernovae), and it has a stronger impact at $z \geq 24$ when Pop. III star formation is dominating. \\

In conclusion, the global evolution of Pop. III stars are strongly dependent on the star formation efficiency because high $\epsilon_{\rm SF, III}$ increases both the SMF and the SFRD. The evolution is also dependent on the IMF adopted (although the IMF affects only the SMF). The critical metallicity has an impact only if we consider a high value, and it mostly affects the self-enrichment as most of the galaxies in our simulation get their metals from their own star formation rather than from an external metal bubble. The critical surface density instead has only a small impact on the Pop. III star formation history and it is restricted to small mass halos and high-$z$. 

\section{Pop. III observability with JWST}
\label{sec:observability}

Unconstrained Pop. III free parameters in our model have an impact on the redshift evolution of the SMF and SFRD. However, our model predicts a consistent number density of Pop. III dominated systems with stellar masses between $10^3 - 10^5$ M$_\odot$. Here we will explore the UV luminosity of these systems using the model described in Section \ref{sec:PopIII_spectra} focusing on the luminosity functions at very high-$z$.\\

\begin{figure*}
    \includegraphics[width=\textwidth]{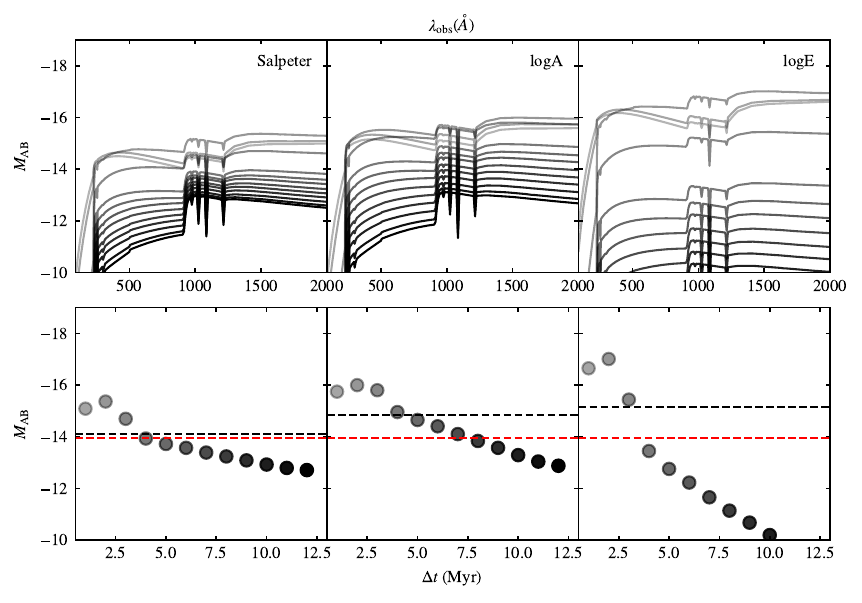}
    \caption{\textbf{top:} time evolution of the SED of the brightest Pop. III galaxy at $z = 8$ assuming a Salpeter (left), logA (mid), logE (right) IMF. Lighter lines assume a smaller $\Delta t$ since the star formation burst. \textbf{bottom:} Absolute UV magnitude of the same galaxy as a function of $\Delta t$ in Myr. The black (red) horizontal line shows the magnitude value of the galaxy assuming a Pop. III (Pop. II) continuous star formation.}
    \label{fig:SalEvoLum}
\end{figure*}

Given the same total stellar mass, Pop. III systems have larger luminosities compared to Pop. II galaxies, but shine for shorter times. This is demonstrated in figure \ref{fig:SalEvoLum} where we show the evolution in the first 10Myr of the SED of a Pop. III galaxy that forms $10^5 M_\odot$ Pop. III stars at z = 8 (top panels) and the absolute UV magnitude (bottom panels) at different times $\Delta t$ since the star formation burst. Left, mid and right panels assume a Pop. III Salpeter, logA and logE IMF respectively. The black horizontal line corresponds to the magnitude computed assuming a continuous star formation throughout the snapshot. For reference, we also show the magnitude of a Pop. II galaxy forming the same amount of Pop. II stars (red dashed line). Firstly, we notice that Pop. III galaxies with continuous star formation and a log-normal IMF are one magnitude brighter than Pop. II galaxies with the same SFR (while there is no significant difference if Pop. III stars follow a Salpeter IMF). Once we focus on the instantaneous star formation model for Pop. III stars, the brightness of the galaxy can be boosted (or reduced) by several magnitudes depending on the time $\Delta t$ following the burst at which we are observing the system (see also \citealt{Trussler2023}).  Even when we consider a Salpeter IMF, the brightness of a Pop. III galaxy changes $\sim$ 2 mags in $\sim$ 10Myr. This change is more dramatic for the log-normal IMFs (especially for the logE IMF where in 10 Myr $\Delta M_{\rm AB} \simeq 7$ mags). 
%The large difference between the instantaneous and continuous star formation models supports our choice of using the instantaneous model (taking a random $\Delta t$ for each Pop. III dominated galaxy) when computing the luminosity of these systems. \\
The large difference between the instantaneous and continuous star formation models shown in Fig. \ref{fig:SalEvoLum} for a single Pop. III galaxy will reflect in the entire population and potentially lead to a higher number density of extremely bright Pop. III galaxies.\\
Fig. \ref{fig:LF16} shows the luminosity function (computed as described in the Appendix \ref{sec:AppendixB}). As a result of instantaneous Pop. III star formation, there are chances to observe these galaxies at a time when their luminosities are higher than what is given by the continuous scenario. This is reflected in Fig. \ref{fig:LF16} where we see the bright end of the solid cyan line shifting to higher luminosities. When looking at the total luminosity function we see a much significant Pop. III contribution to the number density of galaxies around M$_{\rm AB} \sim -16$ when assuming a LogE IMF for their SF. \\
%Because Pop. III star formation occurs in a single burst, the solid cyan line shifts to higher luminosities. 
%However, when looking at the total luminosity function, except for the bright end of our distribution ($M_{\rm UV} < -15$) with the logE IMF (lower panel), the results assume Pop. III SF is continuous (solid line) or in a burst (dashed) are almost identical. This is because the Pop. III dominated systems are quite rare to observe compared to the more long-lived and more numerous Pop. II galaxies. Nevertheless, this result shows that for a more top-heavy IMF, at least some of the brightest systems at very high-$z$ are likely to be Pop. III dominated.\\

\begin{figure}
    \includegraphics[width=\columnwidth]{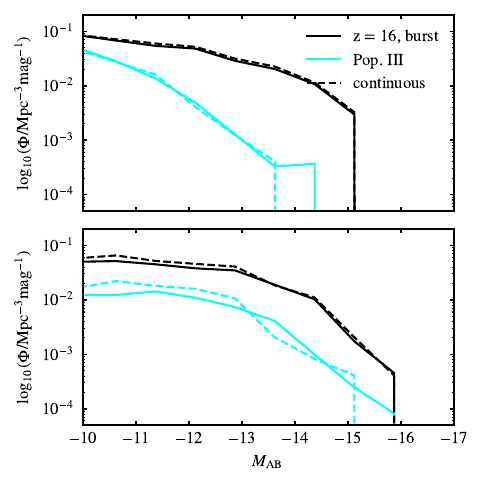}
    \caption{Total (black) and Pop.III (cyan) UV luminosity function at $z = 16$ assuming the Salpeter (top) and the logE (bottom) IMF. Solid lines are computed assuming that Pop. III stars form in a single burst, dashed lines take a continuous star formation model. }
    \label{fig:LF16}
\end{figure}

\begin{figure*}
    \includegraphics[width=\textwidth]{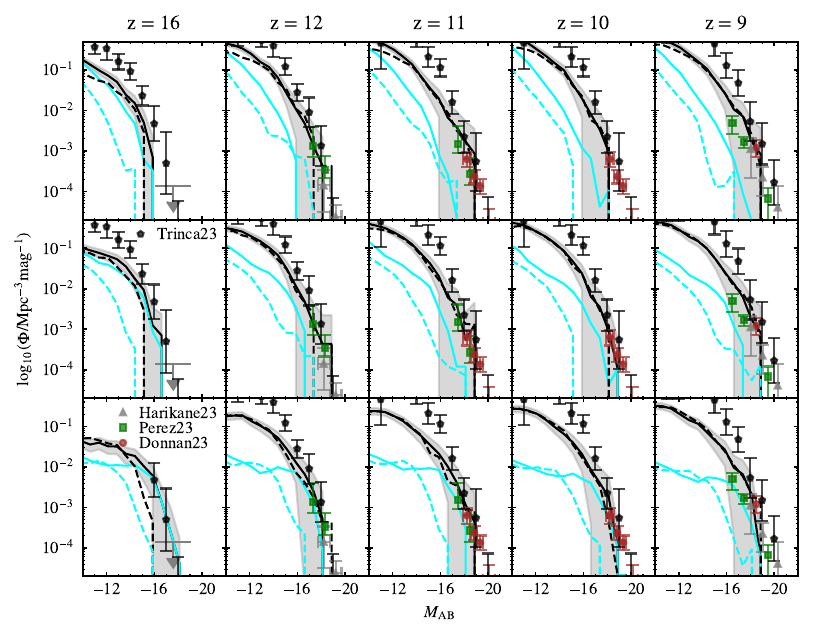}
    \caption{Total (black) and Pop. III dominated galaxies (cyan) UV LF at $z = 16, 12, 11, 10, 9$ (from left to right) for the Salpeter, logA and logE IMF (from top to bottom) and $\alpha_{\rm SF, III} = 0.08$ (solid), 0.008 (dashed). Uncertainties for the black solid line are shown with shaded regions (computed using the Poisson error). Data points highlight recent JWST results from \citet{Donnan2023, Harikane2023, Perez2023} and the results from the simulation of \citet{Trinca2023}.}
    \label{fig:LFdef}
\end{figure*}

Given their larger brightness, a population of Pop. III galaxies with a top-heavy IMF has been suggested as a possible explanation for the abundance of bright galaxies at $z > 10$ (see, e.g. \citealt{Trinca2023, Yung2023} and \citealt{Harikane2023}). 
%This solution seems viable given that, as we saw in Section \ref{sec:global}, the large majority of Pop. III stars form in mini-halos, hence they dominate the global SFRD until $z = 20 -25$ while they become significantly subdominant at $z < 10$. 
Without invoking any exotic physics or a revision of the standard $\Lambda$CDM model of cosmology, other possible explanations are a combination of increased star formation efficiencies at high-$z$ and reduced feedback (\citealt{Qin2023}), bursty star formation (\citealt{Sun2023}), cosmic variance (\citealt{Shen2023}) and a modified $\Lambda$CDM power spectrum (\citealt{Parashari2023,Padmanabhan2023}). Here, we show the predicted UV luminosity functions at $z = 16, 12, 11, 10, 9$ (we chose these values as the ones for which there are more JWST observations available). Results for different IMFs and different star formation efficiencies are summarized in Fig. \ref{fig:LFdef} where all the other Pop. III parameters are taken as in the fiducial model. For all the models considered below, our total luminosity function agrees with early JWST observations at $z \leq 12$ (except for the points at $M_{\rm AB} < -19$ where we do not have galaxies in \textsc{meraxes}). 

With our fiducial parameters (dashed lines), both with the Salpeter and with the logA IMF (upper and middle row), we find that Pop. III systems are not the brightest galaxies at the redshift considered (M$_{\rm AB} \geq -14$). However, when we consider a log-normal IMF with a characteristic mass of 60M$_\odot$ (lower row), Pop. III galaxies become significantly brighter (the cyan line approaches the black line) and at $z = 12$ and 16 when some of the brightest systems (M$_{\rm AB} \sim -16$) are Pop. III dominated. Even though most of the Pop. III galaxies are still very faint (well below the sensitivity of JWST). This result may indicate that Pop. III dominated galaxies are present at $z > 12$. \\
%This result is even more surprising considering that in this model the star formation efficiency of Pop. III stars is one order of magnitude below the Pop. II one. 
%Given the small size of the simulation we intrinsically miss the brightest galaxies and so we cannot give any strong conclusions, 
This result becomes more robust when considering $\alpha_{\rm SF, III} = \alpha_{\rm SF, II}$ (solid lines). The bright end of the total and Pop. III dominated galaxy UVLF shifts of $1-2$ magnitudes (depending on the IMF) at $z = 16$ and 12. At $z \leq 11$, there is no or little difference in the total UVLF with different star formation efficiency even though Pop. III dominated galaxies are still $\sim 1$ mag brighter. Models with a Salpeter or a log-normal IMF centered at $10 M_\odot$ predict that the bright end of the UVLF at $z = 16$ is impacted by Pop. III dominated systems while at $z \leq 12$ none or very few of the brightest galaxies are Pop. III dominated. Overall, the abundance of bright galaxies at $z \geq 12$ is better explained by the model with the logE Pop. III IMF and $\alpha_{\rm SF, III} = \alpha_{\rm SF, II} = 0.08$. For this model at $z \geq 12$ we find Pop. III dominated galaxies with M$_{\rm AB} \simeq -18$ and at $z = 16$ all the brightest galaxies (M$_{\rm AB} \leq -16$) are Pop. III dominated.\\
%At $z < 12$, bright Pop. III dominated systems become less common, and the bright end of the UVLF (M$_{\rm AB} \simeq -19$) is not impacted by these systems. \\
%Add here comparison with Trinca+23
The results discussed above considered a star formation efficiency $\alpha_{\rm SF, III} \geq 0.008$ which allows the formation of Pop. III stellar systems in \textsc{meraxes} with masses up to $\sim 10^5 M_\odot$. These values are substantially higher than what predicted by some hydro-dynamical simulations \citep[e.g.][]{Xu2016,Sarmento2022}{}{}. This may suggest lower star formation efficiencies than what considered in this work making Pop. III galaxies much fainter and very unlikely to be observed.
We compared our results with \citet{Trinca2023} (black pentagons in Fig. \ref{fig:LFdef}), and found a good agreement at $z \leq 12$ and M$_{\rm AB} \lesssim -16.5$ for all the models considered. The main differences are that at $z = 9$, we do not have any galaxy with M$_{\rm AB} \sim -20$, and their model has a steeper evolution predicting many more faint galaxies. At $z = 16$ only models with $\alpha_{\rm SF, III} = 0.08$ and a log-normal IMF reproduce their results at $-14 \leq$ M$_{\rm AB} \leq -18$. Even though the total UVLF are similar, their Pop. III contribution is significantly lower, especially when compared to the models with $\alpha_{\rm SF, III} = 0.08$ (their results predict that less than 10\% of bright galaxies at $z = 15-16$ host an active Pop. III stellar population). Given that the choice of Pop. III free parameters are similar (they also take $\alpha_{\rm SF, III} \sim 0.1$ and account for a critical surface density of the cold gas before forming Pop. III stars). The main differences in the model that can explain the different results are \textit{(i)} the homogeneous feedback and \textit{(ii)} the Pop. III luminosity calculation. In their model, both the radiative feedback and the chemical enrichment are homogeneous; this might overestimate the suppression of Pop. III systems at $z \sim 10-20$ (when the LW background is effective). Regarding the second point, our model accounts for the fact that a Pop. III galaxies might be observed immediately after they formed stars. As shown in Fig. \ref{fig:LF16}, this enables us to boost the luminosity of some Pop. III dominated galaxies by several magnitudes. Without accounting for this effect, our Pop. III galaxies would have similar luminosity to those predicted by \citet{Trinca2023}.\\    
The main limitation of our results is that, given the small size of the simulation, we intrinsically miss the brightest galaxies (at $z \geq 9$ we do not get galaxies brighter than M$_{\rm AB} \simeq -19$). Larger simulation volumes will allow us to be conclusive, but this result suggests that if Pop. III stars are still forming at $z \geq 10-12$, and they have a top-heavy IMF, they do not require increased star formation efficiencies to outshine the Pop. II systems. 
As Pop. III dominated galaxies become much more rare at $z \lesssim 10$, this scenario allows us to increase the abundance of bright galaxies only at $z > 10$ without boosting the bright end of the UV luminosity function at lower redshift hence we achieve a good agreement with early JWST observations from \citet{Donnan2023, Harikane2023} and \citet{Perez2023}.  

\section{Conclusions}
\label{sec:conclusions}

In this paper, we studied Pop. III star formation in mini-halos with an updated version of the \textsc{meraxes} semi-analytic model of galaxy formation that includes Lyman-Werner background and the streaming velocities. We also implemented external metal enrichment following the growth of the supernova bubbles according to the Sedov-Taylor model. We computed the bubble size distribution function and found that our results agree with \citet{Trenti2009} (most bubbles are smaller than $150 h^{-1}$ ckpc at $z = 6$ and have a typical size of $100-200$ ckpc at $z = 5$) hence most halos get self-enriched rather than externally polluted. Only low-mass halos (M$_{\rm vir} \leq 10^{7.5} M_\odot$) at $z < 10$ are more likely to get their metals from the IGM. This is a consequence of the small size of the supernova bubbles that results in a small filling factor (0.1\% at $z = 12$ and less than 1\% at $z = 5$.)\\ 
The free parameters allow us to explore the global properties of Pop. III dominated galaxies. We ran this model on top of a dark matter-only N-body simulation able to resolve all the mini-halos down to $\sim 3 \times 10^5 M_\odot$ at $z \leq 30$. We explored the impact on the SMF and SFRD of the main free parameters of our model. All models converge at $z \lesssim 10$. However, Pop. III star formation efficiency and IMF lead to differences at high-$z$.\\
%The model allows us to compute the luminosity of Pop. III galaxies (assuming either instantaneous or continuous star formation) given the SED of the Pop. III stellar population. 
Finally, we investigated the SED evolution of a Pop. III dominated galaxy for different IMFs. The shorter lifetime of a Pop. III galaxy motivated us to use an instantaneous star formation model when computing the luminosity function. We computed the total and the Pop. III galaxy UV luminosity functions and compared these to the early JWST results in order to study whether the excess of bright galaxies at high-$z$ could be explained by a population of Pop. III dominated galaxies. Having explored different IMFs and star formation efficiencies, our model predicts, for a log-normal IMF with a characteristic mass of 60 M$_\odot$ and a Pop. III star formation efficiency comparable to the Pop. II, most of the brightest galaxies at $z \geq 12$ and all at $z = 16$ (M$_{\rm AB} = - 18$) are Pop. III dominated. Even with a smaller mass or a Salpeter IMF, Pop. III dominated systems still have an impact on the bright end of the UVLF at $z = 16$. 
%with a top-heavy IMF with a star formation efficiency equal to their Pop. II counterparts. For a log-normal IMF with a characteristic mass of 60 M$_\odot$ and a Pop. III star formation efficiency $\alpha_{\rm SF, III} = \alpha_{\rm SF, II} = 0.08$ most of the brightest galaxies at $z \geq 12$ and all at $z = 16$ (M$_{\rm AB} = - 18$) are Pop. III dominated. Adopting a smaller characteristic mass or a Salpeter IMF, Pop. III dominated systems still have an impact on the bright end of the UVLF, but only at $z = 16$. 
%If instead, we adopt $\alpha_{\rm SF, III} = 0.008$, it becomes extremely difficult to obtain Pop. III systems with luminosities comparable (or larger) than the Pop. II, even with the most top-heavy IMF. 
In conclusion, this work supports the scenario for which top-heavy Pop. III dominated galaxies might explain the abundance of bright JWST galaxies at $z \geq 12$ without requiring very high star formation efficiencies or extremely weak feedback at high-$z$. \\ 

\section*{Acknowledgements}

We thank the anonymous referee for their insightful comments which helped us improve this work. This research was supported by the Australian Research Council Centre of Excellence for All Sky Astrophysics in 3 Dimensions (ASTRO 3D, project \#CE170100013). This work was performed on the OzSTAR national facility at Swinburne University of Technology. OzSTAR is funded by Swinburne University of Technology and the National Collaborative Research Infrastructure Strategy (NCRIS). Finally, we thank D. Schaerer for sharing with us the Pop. III SEDs and for the insightful discussions.

%%%%%%%%%%%%%%%%%%%%%%%%%%%%%%%%%%%%%%%%%%%%%%%%%%
\section*{Data Availability}

%The data underlying this article will be shared on reasonable request to the corresponding author.
The main data presented in this work have been analysed with the \textsc{dragons}\footnote{\url{https://github.com/meraxes-devs/dragons}} and \textsc{sector}\footnote{\url{https://github.com/meraxes-devs/sector}} python packages publicly available. The latest version of \textsc{meraxes} is now publicly available on GitHub here\footnote{\url{https://github.com/meraxes-devs/meraxes}}. The output of the high-resolution N-body simulation used in this work can be downloaded from Zenodo at \citet{balu2024}. Further data and codes will be shared on reasonable request to the corresponding author.



\bibliographystyle{mnras}
\bibliography{meraxes} % if your bibtex file is called meraxes.bib

\begin{thebibliography}{}
\makeatletter
\relax
\def\mn@urlcharsother{\let\do\@makeother \do\$\do\&\do\#\do\^\do\_\do\%\do\~}
\def\mn@doi{\begingroup\mn@urlcharsother \@ifnextchar [ {\mn@doi@}
  {\mn@doi@[]}}
\def\mn@doi@[#1]#2{\def\@tempa{#1}\ifx\@tempa\@empty \href
  {http://dx.doi.org/#2} {doi:#2}\else \href {http://dx.doi.org/#2} {#1}\fi
  \endgroup}
\def\mn@eprint#1#2{\mn@eprint@#1:#2::\@nil}
\def\mn@eprint@arXiv#1{\href {http://arxiv.org/abs/#1} {{\tt arXiv:#1}}}
\def\mn@eprint@dblp#1{\href {http://dblp.uni-trier.de/rec/bibtex/#1.xml}
  {dblp:#1}}
\def\mn@eprint@#1:#2:#3:#4\@nil{\def\@tempa {#1}\def\@tempb {#2}\def\@tempc
  {#3}\ifx \@tempc \@empty \let \@tempc \@tempb \let \@tempb \@tempa \fi \ifx
  \@tempb \@empty \def\@tempb {arXiv}\fi \@ifundefined
  {mn@eprint@\@tempb}{\@tempb:\@tempc}{\expandafter \expandafter \csname
  mn@eprint@\@tempb\endcsname \expandafter{\@tempc}}}

\bibitem[\protect\citeauthoryear{{Balu}, {Greig}, {Qiu}, {Power}, {Qin},
  {Mutch}  \& {Wyithe}}{{Balu} et~al.}{2023}]{Balu2023}
{Balu} S.,  {Greig} B.,  {Qiu} Y.,  {Power} C.,  {Qin} Y.,  {Mutch} S.,
  {Wyithe} J. S.~B.,  2023, \mn@doi [\mnras] {10.1093/mnras/stad281}, \href
  {https://ui.adsabs.harvard.edu/abs/2023MNRAS.520.3368B} {520, 3368}

\bibitem[\protect\citeauthoryear{{Balu}, {Power}, {Mutch}, {Qin}  \&
  {Ventura}}{{Balu} et~al.}{2024}]{balu2024}
{Balu} S.,  {Power} C.,  {Mutch} S.,  {Qin} Y.,   {Ventura} E.~M.,  2024,
  \mn@doi{10.5281/zenodo.10608236}, \url
  {https://doi.org/10.5281/zenodo.10608236}

\bibitem[\protect\citeauthoryear{{Barkana} \& {Loeb}}{{Barkana} \&
  {Loeb}}{2001}]{Barkana2001}
{Barkana} R.,  {Loeb} A.,  2001, \mn@doi [\physrep]
  {10.1016/S0370-1573(01)00019-9}, \href
  {https://ui.adsabs.harvard.edu/abs/2001PhR...349..125B} {349, 125}

\bibitem[\protect\citeauthoryear{{Barkana} \& {Loeb}}{{Barkana} \&
  {Loeb}}{2005}]{Barkana2005}
{Barkana} R.,  {Loeb} A.,  2005, \mn@doi [\apj] {10.1086/429954}, \href
  {https://ui.adsabs.harvard.edu/abs/2005ApJ...626....1B} {626, 1}

\bibitem[\protect\citeauthoryear{{Bosman} et~al.,}{{Bosman}
  et~al.}{2022}]{Bosman2022}
{Bosman} S. E.~I.,  et~al., 2022, \mn@doi [\mnras] {10.1093/mnras/stac1046},
  \href {https://ui.adsabs.harvard.edu/abs/2022MNRAS.514...55B} {514, 55}

\bibitem[\protect\citeauthoryear{{Bromm}, {Coppi}  \& {Larson}}{{Bromm}
  et~al.}{1999}]{Bromm1999}
{Bromm} V.,  {Coppi} P.~S.,   {Larson} R.~B.,  1999, \mn@doi [\apjl]
  {10.1086/312385}, \href
  {https://ui.adsabs.harvard.edu/abs/1999ApJ...527L...5B} {527, L5}

\bibitem[\protect\citeauthoryear{{Chiaki} \& {Wise}}{{Chiaki} \&
  {Wise}}{2023}]{Chiaki2023}
{Chiaki} G.,  {Wise} J.~H.,  2023, \mn@doi [\mnras] {10.1093/mnras/stad433},
  \href {https://ui.adsabs.harvard.edu/abs/2023MNRAS.520.5077C} {520, 5077}

\bibitem[\protect\citeauthoryear{{Chiaki} \& {Yoshida}}{{Chiaki} \&
  {Yoshida}}{2022}]{Chiaki2022}
{Chiaki} G.,  {Yoshida} N.,  2022, \mn@doi [\mnras] {10.1093/mnras/stab2799},
  \href {https://ui.adsabs.harvard.edu/abs/2022MNRAS.510.5199C} {510, 5199}

\bibitem[\protect\citeauthoryear{{Chon}, {Omukai}  \& {Schneider}}{{Chon}
  et~al.}{2021}]{Chon2021}
{Chon} S.,  {Omukai} K.,   {Schneider} R.,  2021, \mn@doi [\mnras]
  {10.1093/mnras/stab2497}, \href
  {https://ui.adsabs.harvard.edu/abs/2021MNRAS.508.4175C} {508, 4175}

\bibitem[\protect\citeauthoryear{{Chon}, {Ono}, {Omukai}  \&
  {Schneider}}{{Chon} et~al.}{2022}]{Chon2022}
{Chon} S.,  {Ono} H.,  {Omukai} K.,   {Schneider} R.,  2022, \mn@doi [\mnras]
  {10.1093/mnras/stac1549}, \href
  {https://ui.adsabs.harvard.edu/abs/2022MNRAS.514.4639C} {514, 4639}

\bibitem[\protect\citeauthoryear{{Correa Magnus}, {Smith}, {Khochfar},
  {O'Shea}, {Wise}, {Norman}  \& {Turk}}{{Correa Magnus}
  et~al.}{2023}]{Magnus2023}
{Correa Magnus} L.,  {Smith} B.~D.,  {Khochfar} S.,  {O'Shea} B.~W.,  {Wise}
  J.~H.,  {Norman} M.~L.,   {Turk} M.~J.,  2023, \mn@doi [arXiv e-prints]
  {10.48550/arXiv.2307.03521}, \href
  {https://ui.adsabs.harvard.edu/abs/2023arXiv230703521C} {p. arXiv:2307.03521}

\bibitem[\protect\citeauthoryear{{Crosby}, {O'Shea}, {Smith}, {Turk}  \&
  {Hahn}}{{Crosby} et~al.}{2013}]{Crosby2013}
{Crosby} B.~D.,  {O'Shea} B.~W.,  {Smith} B.~D.,  {Turk} M.~J.,   {Hahn} O.,
  2013, \mn@doi [\apj] {10.1088/0004-637X/773/2/108}, \href
  {https://ui.adsabs.harvard.edu/abs/2013ApJ...773..108C} {773, 108}

\bibitem[\protect\citeauthoryear{{Dijkstra}, {Ferrara}  \&
  {Mesinger}}{{Dijkstra} et~al.}{2014}]{Dijkstra2014}
{Dijkstra} M.,  {Ferrara} A.,   {Mesinger} A.,  2014, \mn@doi [\mnras]
  {10.1093/mnras/stu1007}, \href
  {https://ui.adsabs.harvard.edu/abs/2014MNRAS.442.2036D} {442, 2036}

\bibitem[\protect\citeauthoryear{{Donnan} et~al.,}{{Donnan}
  et~al.}{2023}]{Donnan2023}
{Donnan} C.~T.,  et~al., 2023, \mn@doi [\mnras] {10.1093/mnras/stac3472}, \href
  {https://ui.adsabs.harvard.edu/abs/2023MNRAS.518.6011D} {518, 6011}

\bibitem[\protect\citeauthoryear{{Elahi}, {Ca{\~n}as}, {Poulton}, {Tobar},
  {Willis}, {Lagos}, {Power}  \& {Robotham}}{{Elahi} et~al.}{2019a}]{VR}
{Elahi} P.~J.,  {Ca{\~n}as} R.,  {Poulton} R. J.~J.,  {Tobar} R.~J.,  {Willis}
  J.~S.,  {Lagos} C. d.~P.,  {Power} C.,   {Robotham} A. S.~G.,  2019a, \mn@doi
  [\pasa] {10.1017/pasa.2019.12}, \href
  {https://ui.adsabs.harvard.edu/abs/2019PASA...36...21E} {36, e021}

\bibitem[\protect\citeauthoryear{{Elahi}, {Poulton}, {Tobar}, {Ca{\~n}as},
  {Lagos}, {Power}  \& {Robotham}}{{Elahi} et~al.}{2019b}]{treefrog}
{Elahi} P.~J.,  {Poulton} R. J.~J.,  {Tobar} R.~J.,  {Ca{\~n}as} R.,  {Lagos}
  C. d.~P.,  {Power} C.,   {Robotham} A. S.~G.,  2019b, \mn@doi [\pasa]
  {10.1017/pasa.2019.18}, \href
  {https://ui.adsabs.harvard.edu/abs/2019PASA...36...28E} {36, e028}

\bibitem[\protect\citeauthoryear{{Feathers}, {Visbal}, {Kulkarni}  \&
  {Hazlett}}{{Feathers} et~al.}{2023}]{Feathers2023}
{Feathers} C.~R.,  {Visbal} E.,  {Kulkarni} M.,   {Hazlett} R.,  2023, \mn@doi
  [arXiv e-prints] {10.48550/arXiv.2306.07371}, \href
  {https://ui.adsabs.harvard.edu/abs/2023arXiv230607371F} {p. arXiv:2306.07371}

\bibitem[\protect\citeauthoryear{{Fialkov}, {Barkana}, {Tseliakhovich}  \&
  {Hirata}}{{Fialkov} et~al.}{2012}]{Fialkov2012}
{Fialkov} A.,  {Barkana} R.,  {Tseliakhovich} D.,   {Hirata} C.~M.,  2012,
  \mn@doi [\mnras] {10.1111/j.1365-2966.2012.21318.x}, \href
  {https://ui.adsabs.harvard.edu/abs/2012MNRAS.424.1335F} {424, 1335}

\bibitem[\protect\citeauthoryear{{Fukushima}, {Yajima}, {Sugimura}, {Hosokawa},
  {Omukai}  \& {Matsumoto}}{{Fukushima} et~al.}{2020}]{Fukushima2020}
{Fukushima} H.,  {Yajima} H.,  {Sugimura} K.,  {Hosokawa} T.,  {Omukai} K.,
  {Matsumoto} T.,  2020, \mn@doi [\mnras] {10.1093/mnras/staa2062}, \href
  {https://ui.adsabs.harvard.edu/abs/2020MNRAS.497.3830F} {497, 3830}

\bibitem[\protect\citeauthoryear{{Furlanetto} \& {Loeb}}{{Furlanetto} \&
  {Loeb}}{2003}]{FL2003}
{Furlanetto} S.~R.,  {Loeb} A.,  2003, \mn@doi [\apj] {10.1086/374045}, \href
  {https://ui.adsabs.harvard.edu/abs/2003ApJ...588...18F} {588, 18}

\bibitem[\protect\citeauthoryear{{Furlanetto}, {Zaldarriaga}  \&
  {Hernquist}}{{Furlanetto} et~al.}{2004}]{Furlanetto2004}
{Furlanetto} S.~R.,  {Zaldarriaga} M.,   {Hernquist} L.,  2004, \mn@doi [\apj]
  {10.1086/423025}, \href
  {https://ui.adsabs.harvard.edu/abs/2004ApJ...613....1F} {613, 1}

\bibitem[\protect\citeauthoryear{{Galli} \& {Palla}}{{Galli} \&
  {Palla}}{1998}]{Galli1998}
{Galli} D.,  {Palla} F.,  1998, \mn@doi [\aap]
  {10.48550/arXiv.astro-ph/9803315}, \href
  {https://ui.adsabs.harvard.edu/abs/1998A&A...335..403G} {335, 403}

\bibitem[\protect\citeauthoryear{{Greif}, {White}, {Klessen}  \&
  {Springel}}{{Greif} et~al.}{2011}]{Greif2011}
{Greif} T.~H.,  {White} S. D.~M.,  {Klessen} R.~S.,   {Springel} V.,  2011,
  \mn@doi [\apj] {10.1088/0004-637X/736/2/147}, \href
  {https://ui.adsabs.harvard.edu/abs/2011ApJ...736..147G} {736, 147}

\bibitem[\protect\citeauthoryear{{Haiman} \& {Bryan}}{{Haiman} \&
  {Bryan}}{2006}]{Haiman2006}
{Haiman} Z.,  {Bryan} G.~L.,  2006, \mn@doi [\apj] {10.1086/506580}, \href
  {https://ui.adsabs.harvard.edu/abs/2006ApJ...650....7H} {650, 7}

\bibitem[\protect\citeauthoryear{{Harikane} et~al.,}{{Harikane}
  et~al.}{2023}]{Harikane2023}
{Harikane} Y.,  et~al., 2023, \mn@doi [\apjs] {10.3847/1538-4365/acaaa9}, \href
  {https://ui.adsabs.harvard.edu/abs/2023ApJS..265....5H} {265, 5}

\bibitem[\protect\citeauthoryear{{Hartwig}, {Glover}, {Klessen}, {Latif}  \&
  {Volonteri}}{{Hartwig} et~al.}{2015}]{Hartwig2015}
{Hartwig} T.,  {Glover} S. C.~O.,  {Klessen} R.~S.,  {Latif} M.~A.,
  {Volonteri} M.,  2015, \mn@doi [\mnras] {10.1093/mnras/stv1368}, \href
  {https://ui.adsabs.harvard.edu/abs/2015MNRAS.452.1233H} {452, 1233}

\bibitem[\protect\citeauthoryear{{Hartwig} et~al.,}{{Hartwig}
  et~al.}{2018}]{Hartwig2018}
{Hartwig} T.,  et~al., 2018, \mn@doi [\mnras] {10.1093/mnras/sty1176}, \href
  {https://ui.adsabs.harvard.edu/abs/2018MNRAS.478.1795H} {478, 1795}

\bibitem[\protect\citeauthoryear{{Hegde} \& {Furlanetto}}{{Hegde} \&
  {Furlanetto}}{2023}]{Hegde2023}
{Hegde} S.,  {Furlanetto} S.~R.,  2023, \mn@doi [\mnras]
  {10.1093/mnras/stad2308}, \href
  {https://ui.adsabs.harvard.edu/abs/2023MNRAS.525..428H} {525, 428}

\bibitem[\protect\citeauthoryear{{Heger} \& {Woosley}}{{Heger} \&
  {Woosley}}{2002}]{Heger2002}
{Heger} A.,  {Woosley} S.~E.,  2002, \mn@doi [\apj] {10.1086/338487}, \href
  {https://ui.adsabs.harvard.edu/abs/2002ApJ...567..532H} {567, 532}

\bibitem[\protect\citeauthoryear{{Heger} \& {Woosley}}{{Heger} \&
  {Woosley}}{2010}]{Heger2010}
{Heger} A.,  {Woosley} S.~E.,  2010, \mn@doi [\apj]
  {10.1088/0004-637X/724/1/341}, \href
  {https://ui.adsabs.harvard.edu/abs/2010ApJ...724..341H} {724, 341}

\bibitem[\protect\citeauthoryear{{Hirano}, {Hosokawa}, {Yoshida}, {Umeda},
  {Omukai}, {Chiaki}  \& {Yorke}}{{Hirano} et~al.}{2014}]{Hirano2014}
{Hirano} S.,  {Hosokawa} T.,  {Yoshida} N.,  {Umeda} H.,  {Omukai} K.,
  {Chiaki} G.,   {Yorke} H.~W.,  2014, \mn@doi [\apj]
  {10.1088/0004-637X/781/2/60}, \href
  {https://ui.adsabs.harvard.edu/abs/2014ApJ...781...60H} {781, 60}

\bibitem[\protect\citeauthoryear{{Jaacks}, {Thompson}, {Finkelstein}  \&
  {Bromm}}{{Jaacks} et~al.}{2018}]{Jaacks2018}
{Jaacks} J.,  {Thompson} R.,  {Finkelstein} S.~L.,   {Bromm} V.,  2018, \mn@doi
  [\mnras] {10.1093/mnras/sty062}, \href
  {https://ui.adsabs.harvard.edu/abs/2018MNRAS.475.4396J} {475, 4396}

\bibitem[\protect\citeauthoryear{{Jaura}, {Glover}, {Wollenberg}, {Klessen},
  {Geen}  \& {Haemmerl{\'e}}}{{Jaura} et~al.}{2022}]{Jaura2022}
{Jaura} O.,  {Glover} S. C.~O.,  {Wollenberg} K. M.~J.,  {Klessen} R.~S.,
  {Geen} S.,   {Haemmerl{\'e}} L.,  2022, \mn@doi [\mnras]
  {10.1093/mnras/stac487}, \href
  {https://ui.adsabs.harvard.edu/abs/2022MNRAS.512..116J} {512, 116}

\bibitem[\protect\citeauthoryear{{Kauffmann}}{{Kauffmann}}{1996}]{Kauffmann1996}
{Kauffmann} G.,  1996, \mn@doi [\mnras] {10.1093/mnras/281.2.475}, \href
  {https://ui.adsabs.harvard.edu/abs/1996MNRAS.281..475K} {281, 475}

\bibitem[\protect\citeauthoryear{{Kennicutt}}{{Kennicutt}}{1998}]{Kennicutt1998}
{Kennicutt} Robert~C. J.,  1998, \mn@doi [\apj] {10.1086/305588}, \href
  {https://ui.adsabs.harvard.edu/abs/1998ApJ...498..541K} {498, 541}

\bibitem[\protect\citeauthoryear{{Kennicutt} \& {De Los Reyes}}{{Kennicutt} \&
  {De Los Reyes}}{2021}]{Kennicutt2021}
{Kennicutt} Robert~C. J.,  {De Los Reyes} M. A.~C.,  2021, \mn@doi [\apj]
  {10.3847/1538-4357/abd3a2}, \href
  {https://ui.adsabs.harvard.edu/abs/2021ApJ...908...61K} {908, 61}

\bibitem[\protect\citeauthoryear{{Klessen} \& {Glover}}{{Klessen} \&
  {Glover}}{2023}]{Klessen2023}
{Klessen} R.~S.,  {Glover} S. C.~O.,  2023, \mn@doi [\araa]
  {10.1146/annurev-astro-071221-053453}, \href
  {https://ui.adsabs.harvard.edu/abs/2023ARA&A..61...65K} {61, 65}

\bibitem[\protect\citeauthoryear{{Kroupa}}{{Kroupa}}{2001}]{Kroupa2001}
{Kroupa} P.,  2001, \mn@doi [\mnras] {10.1046/j.1365-8711.2001.04022.x}, \href
  {https://ui.adsabs.harvard.edu/abs/2001MNRAS.322..231K} {322, 231}

\bibitem[\protect\citeauthoryear{{Kulkarni}, {Visbal}  \& {Bryan}}{{Kulkarni}
  et~al.}{2021}]{Kulkarni2021}
{Kulkarni} M.,  {Visbal} E.,   {Bryan} G.~L.,  2021, \mn@doi [\apj]
  {10.3847/1538-4357/ac08a3}, \href
  {https://ui.adsabs.harvard.edu/abs/2021ApJ...917...40K} {917, 40}

\bibitem[\protect\citeauthoryear{{Magg} et~al.,}{{Magg}
  et~al.}{2022}]{Magg2022}
{Magg} M.,  et~al., 2022, \mn@doi [\mnras] {10.1093/mnras/stac1664}, \href
  {https://ui.adsabs.harvard.edu/abs/2022MNRAS.514.4433M} {514, 4433}

\bibitem[\protect\citeauthoryear{{Maiolino} et~al.,}{{Maiolino}
  et~al.}{2023}]{Maiolino2023}
{Maiolino} R.,  et~al., 2023, \mn@doi [arXiv e-prints]
  {10.48550/arXiv.2306.00953}, \href
  {https://ui.adsabs.harvard.edu/abs/2023arXiv230600953M} {p. arXiv:2306.00953}

\bibitem[\protect\citeauthoryear{{Mebane}, {Mirocha}  \& {Furlanetto}}{{Mebane}
  et~al.}{2018}]{Mebane2018}
{Mebane} R.~H.,  {Mirocha} J.,   {Furlanetto} S.~R.,  2018, \mn@doi [\mnras]
  {10.1093/mnras/sty1833}, \href
  {https://ui.adsabs.harvard.edu/abs/2018MNRAS.479.4544M} {479, 4544}

\bibitem[\protect\citeauthoryear{{Mesinger}, {Furlanetto}  \& {Cen}}{{Mesinger}
  et~al.}{2011}]{Mesinger2011}
{Mesinger} A.,  {Furlanetto} S.,   {Cen} R.,  2011, \mn@doi [\mnras]
  {10.1111/j.1365-2966.2010.17731.x}, \href
  {https://ui.adsabs.harvard.edu/abs/2011MNRAS.411..955M} {411, 955}

\bibitem[\protect\citeauthoryear{{Mutch}, {Geil}, {Poole}, {Angel}, {Duffy},
  {Mesinger}  \& {Wyithe}}{{Mutch} et~al.}{2016}]{Mutch2016}
{Mutch} S.~J.,  {Geil} P.~M.,  {Poole} G.~B.,  {Angel} P.~W.,  {Duffy} A.~R.,
  {Mesinger} A.,   {Wyithe} J. S.~B.,  2016, \mn@doi [\mnras]
  {10.1093/mnras/stw1506}, \href
  {https://ui.adsabs.harvard.edu/abs/2016MNRAS.462..250M} {462, 250}

\bibitem[\protect\citeauthoryear{{Mutch}, {Greig}, {Qin}, {Poole}  \&
  {Wyithe}}{{Mutch} et~al.}{2023}]{Mutch2023}
{Mutch} S.~J.,  {Greig} B.,  {Qin} Y.,  {Poole} G.~B.,   {Wyithe} J. S.~B.,
  2023, \mn@doi [arXiv e-prints] {10.48550/arXiv.2303.07378}, \href
  {https://ui.adsabs.harvard.edu/abs/2023arXiv230307378M} {p. arXiv:2303.07378}

\bibitem[\protect\citeauthoryear{{Nebrin}, {Giri}  \& {Mellema}}{{Nebrin}
  et~al.}{2023}]{Nebrin2023}
{Nebrin} O.,  {Giri} S.~K.,   {Mellema} G.,  2023, \mn@doi [\mnras]
  {10.1093/mnras/stad1852}, \href
  {https://ui.adsabs.harvard.edu/abs/2023MNRAS.524.2290N} {524, 2290}

\bibitem[\protect\citeauthoryear{{Padmanabhan} \& {Loeb}}{{Padmanabhan} \&
  {Loeb}}{2023}]{Padmanabhan2023}
{Padmanabhan} H.,  {Loeb} A.,  2023, \mn@doi [\apjl]
  {10.3847/2041-8213/acea7a}, \href
  {https://ui.adsabs.harvard.edu/abs/2023ApJ...953L...4P} {953, L4}

\bibitem[\protect\citeauthoryear{{Pallottini}, {Ferrara}, {Gallerani},
  {Salvadori}  \& {D'Odorico}}{{Pallottini} et~al.}{2014}]{Pallottini2014}
{Pallottini} A.,  {Ferrara} A.,  {Gallerani} S.,  {Salvadori} S.,   {D'Odorico}
  V.,  2014, \mn@doi [\mnras] {10.1093/mnras/stu451}, \href
  {https://ui.adsabs.harvard.edu/abs/2014MNRAS.440.2498P} {440, 2498}

\bibitem[\protect\citeauthoryear{{Parashari} \& {Laha}}{{Parashari} \&
  {Laha}}{2023}]{Parashari2023}
{Parashari} P.,  {Laha} R.,  2023, \mn@doi [\mnras] {10.1093/mnrasl/slad107},
  \href {https://ui.adsabs.harvard.edu/abs/2023MNRAS.526L..63P} {526, L63}

\bibitem[\protect\citeauthoryear{{P{\'e}rez-Gonz{\'a}lez}
  et~al.,}{{P{\'e}rez-Gonz{\'a}lez} et~al.}{2023}]{Perez2023}
{P{\'e}rez-Gonz{\'a}lez} P.~G.,  et~al., 2023, \mn@doi [\apjl]
  {10.3847/2041-8213/acd9d0}, \href
  {https://ui.adsabs.harvard.edu/abs/2023ApJ...951L...1P} {951, L1}

\bibitem[\protect\citeauthoryear{{Prole}, {Clark}, {Klessen}  \&
  {Glover}}{{Prole} et~al.}{2022}]{Prole2022}
{Prole} L.~R.,  {Clark} P.~C.,  {Klessen} R.~S.,   {Glover} S. C.~O.,  2022,
  \mn@doi [\mnras] {10.1093/mnras/stab3697}, \href
  {https://ui.adsabs.harvard.edu/abs/2022MNRAS.510.4019P} {510, 4019}

\bibitem[\protect\citeauthoryear{{Qin} et~al.,}{{Qin} et~al.}{2017}]{Qin2017}
{Qin} Y.,  et~al., 2017, \mn@doi [\mnras] {10.1093/mnras/stx1909}, \href
  {https://ui.adsabs.harvard.edu/abs/2017MNRAS.472.2009Q} {472, 2009}

\bibitem[\protect\citeauthoryear{{Qin}, {Mesinger}, {Park}, {Greig}  \&
  {Mu{\~n}oz}}{{Qin} et~al.}{2020}]{Qin2020}
{Qin} Y.,  {Mesinger} A.,  {Park} J.,  {Greig} B.,   {Mu{\~n}oz} J.~B.,  2020,
  \mn@doi [\mnras] {10.1093/mnras/staa1131}, \href
  {https://ui.adsabs.harvard.edu/abs/2020MNRAS.495..123Q} {495, 123}

\bibitem[\protect\citeauthoryear{{Qin}, {Mesinger}, {Bosman}  \& {Viel}}{{Qin}
  et~al.}{2021}]{Qin2021}
{Qin} Y.,  {Mesinger} A.,  {Bosman} S. E.~I.,   {Viel} M.,  2021, \mn@doi
  [\mnras] {10.1093/mnras/stab1833}, \href
  {https://ui.adsabs.harvard.edu/abs/2021MNRAS.506.2390Q} {506, 2390}

\bibitem[\protect\citeauthoryear{{Qin}, {Balu}  \& {Wyithe}}{{Qin}
  et~al.}{2023}]{Qin2023}
{Qin} Y.,  {Balu} S.,   {Wyithe} J. S.~B.,  2023, \mn@doi [\mnras]
  {10.1093/mnras/stad2448}, \href
  {https://ui.adsabs.harvard.edu/abs/2023MNRAS.526.1324Q} {526, 1324}

\bibitem[\protect\citeauthoryear{{Qiu}, {Mutch}, {da Cunha}, {Poole}  \&
  {Wyithe}}{{Qiu} et~al.}{2019}]{Qiu2019}
{Qiu} Y.,  {Mutch} S.~J.,  {da Cunha} E.,  {Poole} G.~B.,   {Wyithe} J. S.~B.,
  2019, \mn@doi [\mnras] {10.1093/mnras/stz2233}, \href
  {https://ui.adsabs.harvard.edu/abs/2019MNRAS.489.1357Q} {489, 1357}

\bibitem[\protect\citeauthoryear{{Raiter}, {Schaerer}  \& {Fosbury}}{{Raiter}
  et~al.}{2010}]{Raiter2010}
{Raiter} A.,  {Schaerer} D.,   {Fosbury} R.~A.~E.,  2010, \mn@doi [\aap]
  {10.1051/0004-6361/201015236}, \href
  {https://ui.adsabs.harvard.edu/abs/2010A&A...523A..64R} {523, A64}

\bibitem[\protect\citeauthoryear{{Sarmento} \& {Scannapieco}}{{Sarmento} \&
  {Scannapieco}}{2022}]{Sarmento2022}
{Sarmento} R.,  {Scannapieco} E.,  2022, \mn@doi [\apj]
  {10.3847/1538-4357/ac815c}, \href
  {https://ui.adsabs.harvard.edu/abs/2022ApJ...935..174S} {935, 174}

\bibitem[\protect\citeauthoryear{{Sarmento}, {Scannapieco}  \&
  {Cohen}}{{Sarmento} et~al.}{2018}]{Sarmento2018}
{Sarmento} R.,  {Scannapieco} E.,   {Cohen} S.,  2018, \mn@doi [\apj]
  {10.3847/1538-4357/aa989a}, \href
  {https://ui.adsabs.harvard.edu/abs/2018ApJ...854...75S} {854, 75}

\bibitem[\protect\citeauthoryear{{Sassano}, {Schneider}, {Valiante},
  {Inayoshi}, {Chon}, {Omukai}, {Mayer}  \& {Capelo}}{{Sassano}
  et~al.}{2021}]{Sassano2021}
{Sassano} F.,  {Schneider} R.,  {Valiante} R.,  {Inayoshi} K.,  {Chon} S.,
  {Omukai} K.,  {Mayer} L.,   {Capelo} P.~R.,  2021, \mn@doi [\mnras]
  {10.1093/mnras/stab1737}, \href
  {https://ui.adsabs.harvard.edu/abs/2021MNRAS.506..613S} {506, 613}

\bibitem[\protect\citeauthoryear{{Schaerer}}{{Schaerer}}{2002}]{Schaerer2002}
{Schaerer} D.,  2002, \mn@doi [\aap] {10.1051/0004-6361:20011619}, \href
  {https://ui.adsabs.harvard.edu/abs/2002A&A...382...28S} {382, 28}

\bibitem[\protect\citeauthoryear{{Schaller}, {Gonnet}, {Draper}, {Chalk},
  {Bower}, {Willis}  \& {Hausammann}}{{Schaller} et~al.}{2018}]{Schaller2018}
{Schaller} M.,  {Gonnet} P.,  {Draper} P.~W.,  {Chalk} A. B.~G.,  {Bower}
  R.~G.,  {Willis} J.,   {Hausammann} L.,  2018, {SWIFT: SPH With
  Inter-dependent Fine-grained Tasking}, Astrophysics Source Code Library,
  record ascl:1805.020 (\mn@eprint {ascl} {1805.020})

\bibitem[\protect\citeauthoryear{{Schauer}, {Glover}, {Klessen}  \&
  {Clark}}{{Schauer} et~al.}{2021}]{Schauer2021}
{Schauer} A. T.~P.,  {Glover} S. C.~O.,  {Klessen} R.~S.,   {Clark} P.,  2021,
  \mn@doi [\mnras] {10.1093/mnras/stab1953}, \href
  {https://ui.adsabs.harvard.edu/abs/2021MNRAS.507.1775S} {507, 1775}

\bibitem[\protect\citeauthoryear{{Schneider}, {Omukai}, {Inoue}  \&
  {Ferrara}}{{Schneider} et~al.}{2006}]{Schneider2006}
{Schneider} R.,  {Omukai} K.,  {Inoue} A.~K.,   {Ferrara} A.,  2006, \mn@doi
  [\mnras] {10.1111/j.1365-2966.2006.10391.x}, \href
  {https://ui.adsabs.harvard.edu/abs/2006MNRAS.369.1437S} {369, 1437}

\bibitem[\protect\citeauthoryear{{Schneider}, {Omukai}, {Bianchi}  \&
  {Valiante}}{{Schneider} et~al.}{2012}]{Schneider2012}
{Schneider} R.,  {Omukai} K.,  {Bianchi} S.,   {Valiante} R.,  2012, \mn@doi
  [\mnras] {10.1111/j.1365-2966.2011.19818.x}, \href
  {https://ui.adsabs.harvard.edu/abs/2012MNRAS.419.1566S} {419, 1566}

\bibitem[\protect\citeauthoryear{{Shen}, {Vogelsberger}, {Boylan-Kolchin},
  {Tacchella}  \& {Kannan}}{{Shen} et~al.}{2023}]{Shen2023}
{Shen} X.,  {Vogelsberger} M.,  {Boylan-Kolchin} M.,  {Tacchella} S.,
  {Kannan} R.,  2023, \mn@doi [\mnras] {10.1093/mnras/stad2508}, \href
  {https://ui.adsabs.harvard.edu/abs/2023MNRAS.525.3254S} {525, 3254}

\bibitem[\protect\citeauthoryear{{Skinner} \& {Wise}}{{Skinner} \&
  {Wise}}{2020}]{Skinner2020}
{Skinner} D.,  {Wise} J.~H.,  2020, \mn@doi [\mnras] {10.1093/mnras/staa139},
  \href {https://ui.adsabs.harvard.edu/abs/2020MNRAS.492.4386S} {492, 4386}

\bibitem[\protect\citeauthoryear{{Smith}, {Wise}, {O'Shea}, {Norman}  \&
  {Khochfar}}{{Smith} et~al.}{2015}]{Smith2015}
{Smith} B.~D.,  {Wise} J.~H.,  {O'Shea} B.~W.,  {Norman} M.~L.,   {Khochfar}
  S.,  2015, \mn@doi [\mnras] {10.1093/mnras/stv1509}, \href
  {https://ui.adsabs.harvard.edu/abs/2015MNRAS.452.2822S} {452, 2822}

\bibitem[\protect\citeauthoryear{{Sobacchi} \& {Mesinger}}{{Sobacchi} \&
  {Mesinger}}{2013}]{Sobacchi2013}
{Sobacchi} E.,  {Mesinger} A.,  2013, \mn@doi [\mnras] {10.1093/mnras/stt693},
  \href {https://ui.adsabs.harvard.edu/abs/2013MNRAS.432.3340S} {432, 3340}

\bibitem[\protect\citeauthoryear{{Sobacchi} \& {Mesinger}}{{Sobacchi} \&
  {Mesinger}}{2014}]{Sobacchi2014}
{Sobacchi} E.,  {Mesinger} A.,  2014, \mn@doi [\mnras] {10.1093/mnras/stu377},
  \href {https://ui.adsabs.harvard.edu/abs/2014MNRAS.440.1662S} {440, 1662}

\bibitem[\protect\citeauthoryear{{Stacy}, {Bromm}  \& {Loeb}}{{Stacy}
  et~al.}{2011}]{Stacy2011}
{Stacy} A.,  {Bromm} V.,   {Loeb} A.,  2011, \mn@doi [\apjl]
  {10.1088/2041-8205/730/1/L1}, \href
  {https://ui.adsabs.harvard.edu/abs/2011ApJ...730L...1S} {730, L1}

\bibitem[\protect\citeauthoryear{{Stacy}, {Bromm}  \& {Lee}}{{Stacy}
  et~al.}{2016}]{Stacy2016}
{Stacy} A.,  {Bromm} V.,   {Lee} A.~T.,  2016, \mn@doi [\mnras]
  {10.1093/mnras/stw1728}, \href
  {https://ui.adsabs.harvard.edu/abs/2016MNRAS.462.1307S} {462, 1307}

\bibitem[\protect\citeauthoryear{{Sun}, {Faucher-Gigu{\`e}re}, {Hayward}  \&
  {Shen}}{{Sun} et~al.}{2023}]{Sun2023}
{Sun} G.,  {Faucher-Gigu{\`e}re} C.-A.,  {Hayward} C.~C.,   {Shen} X.,  2023,
  \mn@doi [\mnras] {10.1093/mnras/stad2902}, \href
  {https://ui.adsabs.harvard.edu/abs/2023MNRAS.526.2665S} {526, 2665}

\bibitem[\protect\citeauthoryear{{Sutherland} \& {Dopita}}{{Sutherland} \&
  {Dopita}}{1993}]{Sutherland1993}
{Sutherland} R.~S.,  {Dopita} M.~A.,  1993, \mn@doi [\apjs] {10.1086/191823},
  \href {https://ui.adsabs.harvard.edu/abs/1993ApJS...88..253S} {88, 253}

\bibitem[\protect\citeauthoryear{{Tegmark}, {Silk}, {Rees}, {Blanchard}, {Abel}
   \& {Palla}}{{Tegmark} et~al.}{1997}]{Tegmark1997}
{Tegmark} M.,  {Silk} J.,  {Rees} M.~J.,  {Blanchard} A.,  {Abel} T.,   {Palla}
  F.,  1997, \mn@doi [\apj] {10.1086/303434}, \href
  {https://ui.adsabs.harvard.edu/abs/1997ApJ...474....1T} {474, 1}

\bibitem[\protect\citeauthoryear{{Trenti}, {Stiavelli}  \& {Shull}}{{Trenti}
  et~al.}{2009}]{Trenti2009}
{Trenti} M.,  {Stiavelli} M.,   {Shull} J.~M.,  2009, \mn@doi [\apj]
  {10.1088/0004-637X/700/2/1672}, \href
  {https://ui.adsabs.harvard.edu/abs/2009ApJ...700.1672T} {700, 1672}

\bibitem[\protect\citeauthoryear{{Trinca}, {Schneider}, {Valiante}, {Graziani},
  {Zappacosta}  \& {Shankar}}{{Trinca} et~al.}{2022}]{Trinca2022}
{Trinca} A.,  {Schneider} R.,  {Valiante} R.,  {Graziani} L.,  {Zappacosta} L.,
    {Shankar} F.,  2022, \mn@doi [\mnras] {10.1093/mnras/stac062}, \href
  {https://ui.adsabs.harvard.edu/abs/2022MNRAS.511..616T} {511, 616}

\bibitem[\protect\citeauthoryear{{Trinca}, {Schneider}, {Valiante}, {Graziani},
  {Ferrotti}, {Omukai}  \& {Chon}}{{Trinca} et~al.}{2023}]{Trinca2023}
{Trinca} A.,  {Schneider} R.,  {Valiante} R.,  {Graziani} L.,  {Ferrotti} A.,
  {Omukai} K.,   {Chon} S.,  2023, \mn@doi [arXiv e-prints]
  {10.48550/arXiv.2305.04944}, \href
  {https://ui.adsabs.harvard.edu/abs/2023arXiv230504944T} {p. arXiv:2305.04944}

\bibitem[\protect\citeauthoryear{{Trussler} et~al.,}{{Trussler}
  et~al.}{2023}]{Trussler2023}
{Trussler} J. A.~A.,  et~al., 2023, \mn@doi [\mnras] {10.1093/mnras/stad2553},
  \href {https://ui.adsabs.harvard.edu/abs/2023MNRAS.525.5328T} {525, 5328}

\bibitem[\protect\citeauthoryear{{Tseliakhovich} \& {Hirata}}{{Tseliakhovich}
  \& {Hirata}}{2010}]{Tseliakhovich2010}
{Tseliakhovich} D.,  {Hirata} C.,  2010, \mn@doi [\prd]
  {10.1103/PhysRevD.82.083520}, \href
  {https://ui.adsabs.harvard.edu/abs/2010PhRvD..82h3520T} {82, 083520}

\bibitem[\protect\citeauthoryear{{Tumlinson}}{{Tumlinson}}{2006}]{Tumlinson2006}
{Tumlinson} J.,  2006, \mn@doi [\apj] {10.1086/500383}, \href
  {https://ui.adsabs.harvard.edu/abs/2006ApJ...641....1T} {641, 1}

\bibitem[\protect\citeauthoryear{{Venditti}, {Graziani}, {Schneider},
  {Pentericci}, {Di Cesare}, {Maio}  \& {Omukai}}{{Venditti}
  et~al.}{2023}]{Venditti2023}
{Venditti} A.,  {Graziani} L.,  {Schneider} R.,  {Pentericci} L.,  {Di Cesare}
  C.,  {Maio} U.,   {Omukai} K.,  2023, \mn@doi [\mnras]
  {10.1093/mnras/stad1201}, \href
  {https://ui.adsabs.harvard.edu/abs/2023MNRAS.522.3809V} {522, 3809}

\bibitem[\protect\citeauthoryear{{Ventura}, {Trinca}, {Schneider}, {Graziani},
  {Valiante}  \& {Wyithe}}{{Ventura} et~al.}{2023}]{Ventura2023}
{Ventura} E.~M.,  {Trinca} A.,  {Schneider} R.,  {Graziani} L.,  {Valiante} R.,
    {Wyithe} J. S.~B.,  2023, \mn@doi [\mnras] {10.1093/mnras/stad237}, \href
  {https://ui.adsabs.harvard.edu/abs/2023MNRAS.520.3609V} {520, 3609}

\bibitem[\protect\citeauthoryear{{Visbal}, {Haiman}  \& {Bryan}}{{Visbal}
  et~al.}{2015}]{Visbal2015}
{Visbal} E.,  {Haiman} Z.,   {Bryan} G.~L.,  2015, \mn@doi [\mnras]
  {10.1093/mnras/stv1941}, \href
  {https://ui.adsabs.harvard.edu/abs/2015MNRAS.453.4456V} {453, 4456}

\bibitem[\protect\citeauthoryear{{Visbal}, {Haiman}  \& {Bryan}}{{Visbal}
  et~al.}{2018}]{Visbal2018}
{Visbal} E.,  {Haiman} Z.,   {Bryan} G.~L.,  2018, \mn@doi [\mnras]
  {10.1093/mnras/sty142}, \href
  {https://ui.adsabs.harvard.edu/abs/2018MNRAS.475.5246V} {475, 5246}

\bibitem[\protect\citeauthoryear{{Visbal}, {Bryan}  \& {Haiman}}{{Visbal}
  et~al.}{2020}]{Visbal2020}
{Visbal} E.,  {Bryan} G.~L.,   {Haiman} Z.,  2020, \mn@doi [\apj]
  {10.3847/1538-4357/ab994e}, \href
  {https://ui.adsabs.harvard.edu/abs/2020ApJ...897...95V} {897, 95}

\bibitem[\protect\citeauthoryear{{Welch} et~al.,}{{Welch}
  et~al.}{2022}]{Welch2022}
{Welch} B.,  et~al., 2022, \mn@doi [\nat] {10.1038/s41586-022-04449-y}, \href
  {https://ui.adsabs.harvard.edu/abs/2022Natur.603..815W} {603, 815}

\bibitem[\protect\citeauthoryear{{Wollenberg}, {Glover}, {Clark}  \&
  {Klessen}}{{Wollenberg} et~al.}{2020}]{Wollenberg2020}
{Wollenberg} K. M.~J.,  {Glover} S. C.~O.,  {Clark} P.~C.,   {Klessen} R.~S.,
  2020, \mn@doi [\mnras] {10.1093/mnras/staa289}, \href
  {https://ui.adsabs.harvard.edu/abs/2020MNRAS.494.1871W} {494, 1871}

\bibitem[\protect\citeauthoryear{{Xu}, {Norman}, {O'Shea}  \& {Wise}}{{Xu}
  et~al.}{2016}]{Xu2016}
{Xu} H.,  {Norman} M.~L.,  {O'Shea} B.~W.,   {Wise} J.~H.,  2016, \mn@doi
  [\apj] {10.3847/0004-637X/823/2/140}, \href
  {https://ui.adsabs.harvard.edu/abs/2016ApJ...823..140X} {823, 140}

\bibitem[\protect\citeauthoryear{{Yajima}, {Abe}, {Fukushima}, {Ono},
  {Harikane}, {Ouchi}, {Hashimoto}  \& {Khochfar}}{{Yajima}
  et~al.}{2023}]{Yajima2023}
{Yajima} H.,  {Abe} M.,  {Fukushima} H.,  {Ono} Y.,  {Harikane} Y.,  {Ouchi}
  M.,  {Hashimoto} T.,   {Khochfar} S.,  2023, \mn@doi [\mnras]
  {10.1093/mnras/stad2497}, \href
  {https://ui.adsabs.harvard.edu/abs/2023MNRAS.525.4832Y} {525, 4832}

\bibitem[\protect\citeauthoryear{{Yamaguchi}, {Furlanetto}  \&
  {Trapp}}{{Yamaguchi} et~al.}{2023}]{Yamaguchi2023}
{Yamaguchi} N.,  {Furlanetto} S.~R.,   {Trapp} A.~C.,  2023, \mn@doi [\mnras]
  {10.1093/mnras/stad315}, \href
  {https://ui.adsabs.harvard.edu/abs/2023MNRAS.520.2922Y} {520, 2922}

\bibitem[\protect\citeauthoryear{{Yung}, {Somerville}, {Finkelstein}, {Wilkins}
   \& {Gardner}}{{Yung} et~al.}{2023}]{Yung2023}
{Yung} L.~Y.~A.,  {Somerville} R.~S.,  {Finkelstein} S.~L.,  {Wilkins} S.~M.,
  {Gardner} J.~P.,  2023, \mn@doi [\mnras] {10.1093/mnras/stad3484}, \href
  {https://ui.adsabs.harvard.edu/abs/2023MNRAS.tmp.3341Y} {}

\makeatother
\end{thebibliography}

%%%%%%%%%%%%%%%%%%%%%%%%%%%%%%%%%%%%%%%%%%%%%%%%%%

%%%%%%%%%%%%%%%%% APPENDICES %%%%%%%%%%%%%%%%%%%%%

\appendix

\section{Testing the external metal enrichment}
\label{sec:AppendixA}

\begin{figure*}
    \centering
    \includegraphics[width=\textwidth]{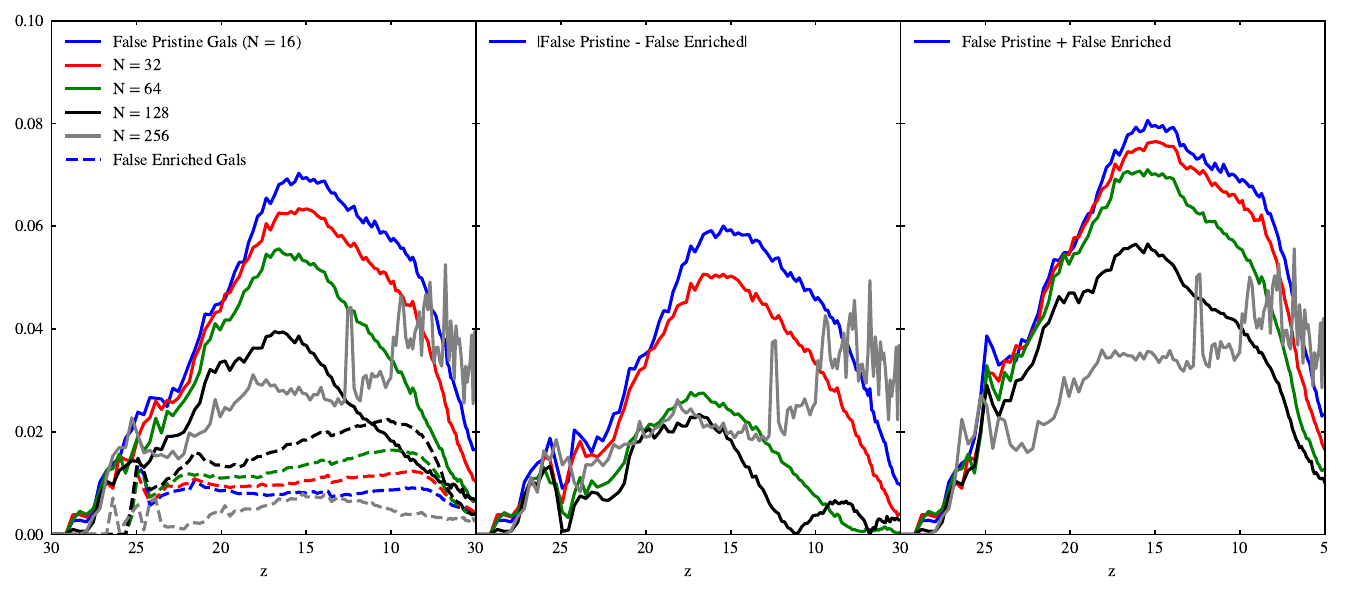}
    \caption{\textbf{left:} mass fraction of false pristine (solid line) and false enriched (dashed line) galaxies (see text for details). Blue, red, green, black and grey lines refer to different grid resolutions (N = 16, 32, 64, 128, 256 respectively).  \textbf{center:} absolute value of the difference between the mass fraction of false pristine and false enriched galaxies. The same colour coding is adopted. \textbf{right:} sum of the mass fraction of false pristine and false enriched galaxies. The same colour coding is adopted.}
    \label{fig:ResTest}
\end{figure*}

In this appendix, we discuss the extent to which our approximation on the external metal enrichment based on the filling factor is appropriate and when it fails (see Section \ref{sec:Metals} for the detailed implementation).\\ 
Since we are avoiding computing all the pairs of distance, there might be some galaxies for which we are not getting a correct enrichment. For this reason, we computed in post-processing the distance between all the pairs of galaxies throughout the entire simulation and we counted how many galaxies in \textsc{meraxes} are marked as externally metal enriched (pristine) even if they are outside (inside) a metal bubble. We repeated this computation for several grid resolutions: N = 16, 32, 64, 128 and 256. Hereafter, we will call "false pristine galaxies" those galaxies that lie inside a metal bubble (and so should be enriched) but in \textsc{meraxes} are labelled as pristine and "false enriched galaxies" those galaxies that are not reached by any bubble but in \textsc{meraxes} are labelled as enriched.\\
In the left panel of Fig. \ref{fig:ResTest} we computed the mass fraction of the false pristine (solid line) and false enriched galaxies (dashed line) using different metal grid resolutions. At lower resolution, the percentage of the false pristine galaxies increases from 4\% at N = 128 (black line) to 7\% at N = 16 (blue line). This happens since, as the pixel becomes bigger, the assumption of the galaxies randomly distributed inside each pixel is no longer valid as the clustering becomes much more important. However, when we use a very fine grid (N = 256, grey line), the mass fraction of false enriched galaxies dramatically increases at $z \leq 10$. This happens because by $z = 10$, the typical bubble size becomes larger than the pixel volume (r$_{\rm bubble} \simeq 40-80$ ckpc, see Fig. \ref{fig:BubbleDF}), and so we lose all the contribution that overflows outside the pixel. The mass percentage of false enriched galaxies instead is only mildly affected by different resolutions (always below 2\%). We then compute the absolute value of the difference between the mass percentage of the false pristine and the false enriched galaxies (middle panel). This quantity tells us if, statistically, we are reproducing the correct global enrichment of the Universe, and it tells us whether or not we can apply this model to study the global properties of the Universe. We can see that having a high-resolution grid significantly improves the quality of our results. For N = 128, this difference is always below 2\%, and at z < 12 is approximately 0\%. This tells us that for a high-resolution grid, globally, we are getting the correct enrichment of the Universe as the mass percentage of false pristine and false enriched galaxies cancels out. This agreement is almost perfect during the EoR, while during the Cosmic Dawn is still very good, but we are underestimating the metal enrichment of $\sim$ 1-2\%. Finally, we are also showing the sum between the false enriched and pristine galaxies (right panel). This quantity instead tells us what is the mass percentage of galaxies for which we are getting a wrong estimation of the enrichment and it is particularly important if we want to apply this model to study analogues. Increasing the resolution improves our result, which, for N = 128, peaks at $\sim$ 4\% during the Cosmic Dawn. In conclusion, our model for N = 128, while reducing the computational cost, still gives an excellent agreement on global enrichment and a very good agreement on the enrichment of single galaxies.\\

%\begin{figure*}
%    \centering
%    \includegraphics[width=\textwidth]{TestEnr_def2.pdf}
%    \caption{As Fig. \ref{fig:ResTest} but accounting for galaxy clustering (see text for details).}
%    \label{fig:ResTest2}
%\end{figure*}

\section{Instantaneous Pop. III star formation}
\label{sec:AppendixB}

In this appendix, we discuss how to evaluate the UV luminosity function of Pop. III galaxies assuming instantaneous star formation. Due to the stochasticity involved in the process of star formation, the burst might occur at any time within a snapshot (with a time duration of $t_{\rm snap}$) for a galaxy, resulting in different luminosities when observed. We refer to $\Delta t$ as the time delay of the star formation burst happening relative to the end of our snapshot and draw its value from a random uniform distribution between zero and $t_{\rm snap}$. One can then evaluate the luminosity of this galaxy, repeat the exercise for all targets after assigning them different $\Delta t$, and calculate their probability distribution as a function of UV magnitude (i.e. the UV luminosity function, $\Phi(M_{\rm UV}, z)$). To achieve a more efficient computation, we instead sample $\Delta t$ in fine steps of $\Delta t_{i} = [0.01, 0.5, 1.0, 1.5, ..., \Delta t_{N}]$ Myr with the last sample bracketing the snapshot length as $\Delta t_{N}\le t_{\rm snap} < \Delta t_{N}+0.5$Myr. Then we calculate the UV magnitudes of each galaxy for given $\Delta t_i$ and evaluate the corresponding luminosity function at the condition of fixed $\Delta t$, $\Phi((M_{\rm UV}, z) | \Delta t_i)$. Finally, the luminosity function is obtained by summing all conditional probability distributions as
\begin{equation}
    \Phi(M_{\rm UV}, z) = \sum_{\Delta t_i = 0.01}^{\Delta t_N} \Phi((M_{\rm UV}, z) | \Delta t_i) \times P(\Delta t_i).
    \label{eq:ProbUV}
\end{equation}
Note that the probability of each conditional luminosity function is simply 
\begin{equation}
P(\Delta t_i) =
\begin{cases}
&\Delta t_i / t_{\rm snap} {\rm\ when\ i=1,}\\
&(\Delta t_i - \Delta t_{i-1})/t_{\rm snap}{\rm\ when\ 1<i<N,}\\
&(\Delta t_i - t_{\rm snap})/t_{\rm snap} {\rm\ when\ i=N}.
\end{cases}
\end{equation}

\bsp	% typesetting comment
\label{lastpage}
\end{document}